**Synthetic spatially graded Rac activation drives directed cell polarization and locomotion**

Benjamin Lin[1], William R. Holmes[2], ChiaoChun Wang[1], Tasuku Ueno[3], Andrew Harwell[1], Leah Edelstein-Keshet[2], Takanari Inoue[3*], Andre Levchenko[1*]

**Abstract**

Migrating cells possess intracellular gradients of Rho GTPases, but it is unknown whether these shallow gradients themselves can induce motility. Here we describe a new method to present cells with induced linear gradients of active, endogenous Rac without receptor activation. Gradients as low as 15% were sufficient to not only trigger cell migration up the synthetic gradient, but also to induce both cell polarization and repolarization. Response kinetics were inversely proportional to Rac gradient values, in agreement with a new mathematical model, suggesting a role for natural input gradient amplification upstream of Rac. Increases in Rac levels beyond a well-defined threshold dramatically augmented polarization and decreased sensitivity to the gradient value. The threshold was governed by initial cell polarity and PI3K activity, supporting a role for both in defining responsiveness to natural or synthetic Rac activation. Our methodology suggests a general way to investigate processes regulated by intracellular signaling gradients.

**Introduction**

Directional motility is an intrinsic ability of many eukaryotic cells to migrate to predetermined locations in an efficient manner. Migration bias is tuned by the detection of various guidance cues that often form spatial gradients. The extracellular gradients of diffusing or surface bound ligands can lead to spatially graded occupancy of extracellular receptors[1]. The spatial asymmetry in receptor occupancy is subsequently translated into an intracellular gradient of polarity effectors which can modify cytoskeleton and lead to the development of an asymmetric cell morphology with functionally distinct front and rear 'compartments'. Remarkably, many cell types can accurately detect less than a 5% difference between ligand concentration at the cell front and back[1-3]. This exquisite sensitivity suggests that intracellular amplification of extracellular cues may be necessary. Indeed, various groups have demonstrated the existence of local positive feedback loops[4-7] as well as mutual inhibition between different regulators as likely candidates for response amplification. However, recent studies have also shown that directional motility can still be achieved, albeit less efficiently, when once thought indispensable molecules involved in putative amplification mechanisms are removed. The complexity of gradient sensing, cell polarization, and taxis responses are thus still actively investigated.

An attractive method for resolving the roles of signaling network components in both spatial cue sensing and directed cell motility is direct activation of these components in a spatially constrained and rapid manner, independent from initiation of upstream, receptor-level signaling. Utilizing this principle, a variety of studies have used optical activation to identify the small Rho GTPase Rac , cofilin[17], thymosin B4[18], and calcium[19] as key components which are sufficient to direct cellular motility. However, an important caveat to these studies has been the reliance upon highly localized activation that can create artificial regional amplification of protein target activity. In contrast, in more physiological settings, a cell processes a shallow gradient of an external cue into a graded intracellular response, as reflected in polarized effectors[20], including those of the small Rho GTPase family[21-23]. Thus it remains unclear if an induced shallow gradient of an active motility signaling component is capable of reconstituting cell polarization and motility. In particular, it is unknown if such perturbations are sufficient to override or enhance endogenous intracellular signaling of the same component. Finally, localized activation of signaling processes presents considerable challenges to quantitative analysis and coupling to detailed computational models developed to describe more natural, spatially distributed signaling events.

To address these questions, we created microfluidic devices permitting generation of precise gradients of extracellular cues[24] and interfaced them with a rapamycin induced dimerization system[25]. In this system, the addition of rapamycin leads to dimerization of two intracellularly transduced molecular components, FK506 binding protein (FKBP) and the rapamycin binding domain of FKBP-rapamycin-binding protein (FRB)[26]. Localization and signaling motifs can be linked to either domain, allowing spatial-temporal control of protein function. We used this system to study the effects of a rapidly induced intracellular gradient of activated Rac, an important regulator of cell polarity[27] and previously shown to induce migration when locally activated.

**Results**

**System Design**

To directly activate endogenous Rac, we introduced two constructs into HeLa cells, a cytoplasm localized effector unit consisting of YFP tagged TIAM1, a Rac GEF, conjugated to FKBP (YF-TIAM1) and an anchor unit at the cell membrane, $Lyn_{11}$-FRB (LDR) . The introduction of rapamycin dimerizes these modified molecular components, thereby bringing TIAM1 in close proximity to the cell membrane where it activates endogenous Rac[25] (Fig. 1a). Due to the chemical nature of the activation, this system is readily amenable to generation of a gradient of Rac activity, accomplished through microfluidics generation of rapamycin gradients. Microfluidic tools have recently been used for the control of complex gradients of extracellular cues, cellular localizations and shaping cell morphologies.

We used a previously developed strategy for imposing diffusion based linear gradients onto cells housed within narrow channels. Specifically, the devices contained a series of 6 μm tall microchannels for cell experimentation, flanked by 130 μm tall main flow-through channels for supplying fresh media and stimulants (see Methods for details on fabrication and Supplementary notes for use, Supplementary Fig. 1a). Actuation of flow led to the development of a linear gradient across the shallow channels due to uneven stimulus concentration in the flow-through channels (Fig. 1b). The microchannels were designed to be on the order of a cell diameter to effectively relegate cells to a uniaxial phenotype and curtail the entry of multiple cells per channel (Fig. 1b). HeLa cells introduced into the microchannels settled into random locations and after a 3-4 hour attachment period were categorized into different states according to the degree and direction of their polarization as evidenced by lamellipodia or lack thereof (Fig.

1c-f). The overall distribution of phenotypes was skewed towards cells with single leading edge lamellipodium, indicating an intrinsic preference for directed motility (Supplementary Fig. 1b). Cells remained viable with limited motility under no flow conditions in the microchannels overnight, supporting complete biocompatibility of the device design (data not shown).

**Direct generation of intracellular Rac gradients**

To explore HeLa cell responsiveness to synthetic generation of an intracellular gradient of Rac activation by induced membrane translocation of TIAM1, we exposed cells to a diffusion mediated linear gradient of rapamycin (0 - 2 nM across the channel or 0.01 nM/μm; yielding front/back concentration differences ranging from ~15-92% across varied cell lengths) visualized by mixing rapamycin solution with Alexa 594, a fluorescent dye with a similar molecular weight (Fig. 1b and Supplementary Fig. 1c,d). We chose this concentration range to avoid saturating the FKBP-FRB system with rapamycin ($K_D$ = 12 nM[34]), which might otherwise have led to the loss of gradeness in cell stimulation. One consequence of using these concentration values was a slow-down of the response kinetics vs. previously reported values for saturating homogeneous rapamycin inputs[25], also observed by us in this system (data not shown), consistent with the reported kinetics of rapamycin mediated FKB-FRB complex formation[34]. The slower response kinetics allowed a better resolution of the effects of gradually accumulating Rac activity, as explored in more detail below. Additionally, the rapamycin concentrations used in our experiments were far below those reported to affect the function of the mammalian target of rapamycin (mTOR)[35]. We tracked both the initial state and subsequent cell responses by imaging over a four hour time period.

Strikingly, we found that the shallow linear gradient of rapamycin could trigger and direct motility of cells in all initial polarity states in the direction up the gradient (Fig. 1c-f and Supplementary Movie 1). Unpolarized (state I) and bipolar cells (state IV) demonstrated symmetry breaking with either the establishment of a leading edge or the enhancement of one lamellipodium and retraction of the other, respectively (Fig. 1d, f and Supplementary Movie 1). Cells already polarized in the direction of the gradient (state II), exhibited widening and extension of the leading lamellipodia and enhanced locomotion (Fig. 1c and Supplementary Movie 1). Surprisingly, cells initially oriented in the direction opposite of the gradient (state III) repolarized and reversed the direction of migration (Fig. 1e and Supplementary Movie 1), a behavior not previously observed in this device geometry[36]. The initiation of cell migration was followed later by a well pronounced enhancement of cell polarity, especially evident when the time lapse data were analyzed in the form of kymographs (Fig. 1g-j) (the behavior addressed in more detail below). The directed cell polarization and migration responses were not observed in untransfected cells (Supplementary Fig. 1e), or if the rapamycin gradient was not imposed (Supplementary Fig. 1f). We also applied our system to MTLn3 cells, a mammary adenocarcinoma line used to assay chemotaxis to epidermal growth factor (EGF) in vitro[37] and in vivo[38] (Supplementary Fig. 2). MTLn3 cells exhibited initial polarity states similar to those seen in Hela and polarized towards gradients of rapamycin (Supplementary Fig. 2). Our data suggests that our system is applicable across multiple cell types and can be applicable to the study of signaling pathways regulating chemotaxis.

To verify that the externally imposed rapamycin gradient was indeed translating into a gradient of Rac activity levels, we used a FRET biosensor for Rac activity (Raichu-Rac)[39], which we first verified in control cells by applying uniform rapamycin stimulation to transfected cells (Supplementary Fig. 3). At a basal level before stimulation, cells demonstrated a graded distribution of active Rac towards the front lamellipodium, as reported previously[22] (Fig. 1k). In state III cells, after gradient introduction, we observed a redistribution of active Rac from the previous front of migrating cells to the side of the cell facing the rapamycin gradient, prior to visible changes in cell polarity, indicating that the rapamycin gradient was indeed translated into an internal gradient of Rac activity (Fig. 1k and Supplementary Movie 2). We monitored the distribution of Mcherry-FKBP-TIAM1 (MF-TIAM1), a construct used to avoid spectral overlap with the FRET reporter, to validate the efficacy of the rapamycin gradient (Fig. 1l). During graded rapamycin stimulation, the difference between the FRET ratio at the front and back of cells continually increased, indicating a graded increase in Rac activity across the cell body (Fig. 1m). Overall our results suggested that an exogenously applied linear gradient of Rac activity was sufficient to direct motility and polarization of cells from a variety of pre-existing polarity phenotypes.

For further validation of our system, we contrasted cellular responses to graded rapamycin with responses to uniform rapamycin stimulation. Cells given a uniform stimulus for the entire experimental period displayed extensive uniform flattening with little net motility (Supplementary Fig. 4a and Supplementary Movie 3). This result was in agreement with previous experiments, showing that differentiated HL-60 cells exposed to spatially uniform stimulation of Rac displayed membrane ruffling around the entire cell periphery[6]. For a more detailed contrast of the effects of graded and uniform intracellular Rac stimulation, we exposed cells to a rapamycin gradient for 2 hours, subsequently followed by uniform stimulation for 3 hours thereafter. As expected, cells polarized and moved in a biased fashion during gradient stimulation, however upon the switch to uniform stimulation, cells rapidly started forming protrusions at the rear side and extended bi-directionally (Supplementary Fig. 4b-c and Supplementary Movie 4). There was a shift of cellular volume from the front to the rear of each cell, as the increased size of the rear cell portion came with a concomitant decrease in the front (Supplementary Fig. 4d). Our results indicate that the loss of directionality in the cue resulted in the propagation of the signal throughout the entire cell and suggests the spatial restriction of Rac activity within a cell is important for maintaining polarization.

**Mathematical modeling of graded Rac inputs**

To better address the simplified, yet non-intuitive, nature of the rapidly-induced graded Rac signaling that bypasses upstream regulation, we developed a mathematical model of cell polarity. The model is based on a simple scheme of Rac-RhoA-Cdc42 small GTPase interactions, expanding on earlier modeling studies (Modeling details found in Supplementary Information). The model made two important qualitative predictions: a) graded input can trigger initial polarization of Rac activity with kinetics inversely related to the gradient strength and b) as the average activity increases, the antagonism between the activities of Rac and Rho GTPases can trigger a phase transition-like sudden increase in polarization, which can be stably maintained as long as the activity of Rac remains high enough (Fig. 2a and Supplementary Information). The model further made explicit predictions with respect to the dependence of the timing of the onset of ensuing polarization responses on two values: 1) the local gradient of Rac activity and 2) the local average value of this activity. In particular, the model predicted that the inverse relationship between the value of the Rac activity gradient and the timing of the initial Rac polarization and initiation of cell migration responses would be accompanied by a much weaker dependence of this timing on the average Rac activity level (Fig. 2b).

**Analysis of Rac gradient values on timing of cellular responses**

To validate the model predictions, we next examined the correlates of the initiation of directed cell migration and their dependencies on the local gradient and average value of rapamycin input. HeLa cell responses were characterized by either the retraction of a leading lamellipodium in cells polarized in the direction opposite that of the gradient (state III), or extension of a pre-existing (state II) or new (state I) lamellipodium in the direction of the gradient (Fig. 2c-e). To quantitatively evaluate these effects, we examined the width and length of the front and rear sides of cells in all states through the entire stimulation period, along with the respective concentrations experienced at each side (Supplementary Fig. 5). Cells that exhibited bipolar phenotypes (state IV) were relatively rare (Fig. 1f and Supplementary Fig. 1b) and therefore were excluded from subsequent analysis. Our analysis revealed that these initial directed migration response times, in agreement with the model predictions, were indeed inversely dependent on rapamycin gradient values across each cell (Fig. 2f-h). This trend was seen across all polarity states (Spearman correlation coefficients; -0.679, -0.655, and -0.583 for state I, II, and III, respectively). Similar dependencies were also observed in MTLn3 cells (Supplementary fig. 6a-c).  The initial response times of state I cells also showed a discernible but much weaker dependency on the local rapamycin concentration, consistent with the model predictions. Cells in states II and III had relatively weaker dependencies (Supplementary Fig. 7a-c, Spearman correlation coefficients; -0.582, -0.478, and -0.377 for state I, II, and III, respectively). State II cells had consistently faster initial response times for any given gradient value when compared to those in state I and state III (Fig. 2f-h and Supplementary Fig. 8) (F test-State II vs. State I, $p < 0.0001$, State II vs. State III, $p < 0.001$), while state I and III cells behaved similarly (State I vs. State III, $p = 0.54$). These results suggest that cells are able to rapidly make polarity decisions in the presence of Rac activity gradients, in congruence with their initial polarization state. Together, our data demonstrates that the magnitude of Rac gradients can influence the timing of the onset of directed cell motility.

Besides the initiation of biased cell migration, the spatially graded activation of Rac eventually triggered a striking, sudden enhancement in cell polarization, with a substantial enlargement of the directed leading lamellipodium (Fig. 3a). This remarkable morphological change was unexpected, but was consistent with the model prediction of an existence of a threshold of Rac activity triggering a rapid, strong and very stable cell polarization, akin to a phase transition. In particular, a gradual variation of the Rac GEF level caused an abrupt transition from a moderately polarized to a highly polarized state, expressed mathematically as a bifurcation in the model response (Fig. 3b). In our experiments cells were exposed to low initial rapamycin concentrations, therefore we further hypothesized that the predicted threshold of strong late cell polarization was due to a gradual accumulation of Rac GEF activity, allowing us to observe the effect of gradual titration of intracellular Rac GEF and Rac activity levels. This gradual Rac GEF build-up was expected to follow the simple mathematical representation of accumulation of Rac GEF concentration over the duration t of early exposure to rapamycin is

$$\Delta[\text{Rac GEF}] = k[\text{Rapamycin}]t, \tag{1}$$

where k is the constant defining tripartite rapamycin-FKBP-FRB complex formation. This expression allowed us to test the hypothesis of the existence of a Rac activity threshold mediating strong polarity symmetry breaking.

One of the immediate consequences of the expansion of the lamellipodium in the direction of the rapamycin gradient was an apparent dimming of the YFP fluorescence signal from the cell body when observed with a wide-field epi-fluroescence microscope (Fig. 3a,c). The dimming of the fluorescence intensity was likely due to a redistribution of cytoplasmic volume from the cell body to the expanding lamellipodium as well as an increase in the translocation of YF-TIAM1 complexes to the membrane (Fig. 3d-e). We thus used the fluorescence intensity of the cell body as the metric of the late polarization onset. This metric indicated that late polarization time exhibited a linear dependency on the mean local rapamycin concentration (Fig. 3f) (Pearson correlation coefficients; -0.608, -0.783, and -0.698 for state I, II, and III, respectively), in agreement with the mathematical model predictions. MTLn3 cells showed similar dependencies on mean rapamycin concentrations (Supplementary Fig. 6d-f). A weaker dependence on the sharpness of the rapamycin gradient, consistent with the model, was also detected in these experiments (Supplementary Fig. 7d-f) (Pearson correlation coefficients; 0.103, -0.511, and -0.228 for state I, II, and III, respectively). As with observations of the early cell responses, state II cells reached the late polarization phase significantly faster than cells in other states for a given rapamycin concentration (Fig. 3f and Supplementary Fig. 9) (ANCOVA test of y intercept, State II vs. State I, p <0.0001, State II vs. State III, p<0.0001) while the difference between the other two states was negligible (State I vs. State III, p=0.13). We also found that a sub-population of state III cells exposed to lower concentrations of rapamycin completely failed to undergo strong polarization (Fig. 3f). In combination, these results suggested the existence of a threshold for late polarization, variable across individual cells and dependent on the initial polarity state.

**Transient graded Rac activation**

The equation (1) further suggested that the duration of rapamycin exposure was as critical as the local rapamycin concentration in exceeding the polarization threshold. Based on *in vitro* estimates of rapamycin-FKBP-FRB complex formation, the characteristic binding time estimated for 1 nM rapamycin is in the range of tens of minutes[34]. To test this prediction, we thus varied the time interval of rapamycin stimulation taking into account the earliest polarization onset time observed for the rapamycin concentrations tested, i.e., 30 min and the estimated equilibration time above. Cells were thus transiently exposed to rapamycin gradients of 0.01 nM/μm for 30 min or 1 hr, followed by perfusion of the devices with rapamycin-free media for the rest of the experiment. We found no cells undergoing late polarization after a 30 minute rapamycin gradient exposure (Fig. 4a-c and Supplementary Movie 5), however, after a 1 hr stimulation, we found subsets of cells in all polarity states able to undergo late polarization, provided that they were exposed to sufficiently high concentrations of rapamycin (Fig. 4d-f and Supplementary Movie 6). The polarization responses were morphologically similar to those seen earlier during continuous stimulation (Fig 4d,e and Supplementary Movie 6), with the timing to the onset of late polarization indistinguishable from those observed for more persistent stimulation (Fig. 4f). Responding cells continued to polarize even after the stimulus was withdrawn, suggesting fixation of the induced polarity and migration states (Fig. 4d-e and Supplementary Movie 6), in contrast to their loss during transition to spatially homogeneous Rac activation (Fig. S3). As mentioned above, this behavior was in agreement with the model predictions of stability of the induced symmetry breaking due to strong antagonism between small GTPases involved, enabling a cell to maintain strongly polarized state whose direction is based on but not continuously informed by the input gradient. Using a classification algorithm to separate responding and non-responding cells, we found that the minimum concentrations needed to elicit a response varied across the polarity states (Fig. 4f). In agreement with the earlier observation of distinct responses in state II cells, cells in this state possessed the lowest response threshold (Fig. 4f) (State I= 1.48 nM, State II = 1.12 nM, State III = 1.44 nM). These data further support the existence of a well-defined, initial polarity-dependent, Rac activation threshold essential for the rapid and profound induced change in polarized cell morphology.

**Inhibition of upstream activators**

Direct activation of Rac allows bypassing of many signaling species commonly thought to be either upstream of Rac or involved in a regulatory feedback with this molecule. A well studied example of such a molecule is phosphatidylinositol 3,4,5-triphosphate ($PIP_3$). In chemoattractant gradients, phosphatidylinositol 3-kinase (PI3K), is recruited to the plasma membrane and phosphorylates the abundant phosphatidylinositol 4,5-bisphosphate ($PIP_2$) to yield $PIP_3$. Due to spatial regulation of PI3K recruitment, $PIP_3$ is often enriched at the front areas of migrating cells[45], displaying an intracellular gradient that is frequently sharper than the gradient of the extracellular chemoattractant. It is thought that PI3K can influence cell guidance through its interaction with small GTPases and actin, but the mechanism of these interactions and the resultant role of PI3K in regulating chemotaxis are still under investigation. If, as sometimes assumed, PI3K is upstream of Rac activation in chemotactic signaling systems, its perturbations are not expected to lead to alteration of cell response to rapamycin-based Rac GEF stimulation. If, on the other hand, PI3K forms a feedback loop with Rac, or otherwise enables Rac-mediated outputs, cell responses might be affected by its inhibition, with the change in cell behavior potentially suggesting the mechanism of PI3K regulatory involvement. Thus, to assess the role of PI3K during Rac-mediated cell polarization and migration we pharmacologically inhibited PI3K with LY294002 at the time of rapamycin stimulation. In contrast to the responses of cells in which PI3K was not perturbed, we observed large subsets of cells exhibiting no response to graded Rac stimulation, both in terms of the initial and late polarization, for all three initial polarity states across various gradients and concentrations (Fig. 5a and Supplementary Fig. 10). Interestingly, the cells that did undergo the initial migration and late polarization responses, did so with the same kinetics as observed in the absence of PI3K perturbation (Fig. 5a). This was consistent with the model prediction that a decrease in simulated strength of the PI3K-mediated feedback to Rac could lead to an increased threshold for cell responsiveness, requiring a sharper effective internal Rac activity gradient for cells to respond (Fig. 5b-c). As a consequence, the stochastic differences in internal states of the cells, defining cell sensitivity to the graded signaling input, can lead to a greater degree of cell population separation into responding and non-responding cells, without affecting the timing of responses in responding cells.

The results in Fig. 3f suggested the existence of a relatively high (re-)polarization threshold for state III cells. Thus we explored whether there would be synergy between this threshold and the increase in polarization threshold caused by inhibition of PI3K. We found that this threshold was indeed shifted in the presence of PI3K inhibition to a higher rapamycin concentration level (Fig. 5d) (0.97 nM for untreated state III vs. 1.41 nM for LY294002 treated state III). These results thus support the notion that PI3K can serve to sensitize cells to spatially graded Rac activation, allowing them to more readily exceed polarization and re-polarization thresholds.

**Discussion**

The results presented in this report argue that directly induced, spatially graded membrane translocation of a Rac activator, TIAM1, can trigger unambiguous polarization and directed movement of cells, aligned with the direction of the stimulation gradient. The gradients of the inducer of Rac activation, the exogenously added rapamycin, can be effective with values as low as 15% across the cell length, with the rates of cellular responses to the stimulation being defined by the gradient value. The results suggest that, in the context of the more natural gradient sensing, in which the cue is present in the form of a chemoattractant ligand, amplification of the input gradient upstream of Rac activation can be relatively mild or non-existent for a cell to adequately respond. However, the kinetics of response can be strongly enhanced if the cell faces sharper input gradients or if these gradients are strongly amplified upstream of Rac activation.

The rapamycin-induced activation of Rac combined with microfluidic control of subtle changes in the input gradient and concentration also enables screening of the effects of slow variation in the total cellular Rac activity. Both a simple model describing a feedback-based interplay between small GTPases in a cell and the corresponding experimental observations support a novel finding that rapid and pronounced transition to a much stronger degree of polarization can occur if Rac activity exceeds a threshold level. This threshold was found to be strongly affected by the initial polarization status of the responding cell. Thus cells that are initially polarized in the direction of the applied gradient on average have lower response thresholds than cells that are unpolarized or polarized in the opposite direction, so that, in contrast to cells in other initial polarization states, only a fraction of cells polarized against the gradient responded to graded Rac stimulation. These results are consistent with the following view supported by the model: an existing endogenous gradient of Rac activity in state II cells combined with the concordant exogenous Rac activity gradient would lead to a smaller difference between the maximum local initial Rac activity within a cell and the polarization threshold value, relative to other cell states, thus sensitizing the state II cells but not cells in other states. More generally, these results support the notion that superposition of external and internal polarization cues affects the direction of cell migration, and furthermore that the relevant decision making occurs at the level of Rac activity stimulation.

Our results further suggest that signal processing upstream of Rac activation in the context of natural chemoattractant stimulation may limit the degree of total Rac activation and thus the ability of the cell to reach the threshold controlling transition into the strongly polarized state. Thus a single ligand may not induce such a transition. However, the threshold might potentially be reached and exceeded given multiple inputs converging on Rac activation, which may be affected by cell type-specific peculiarities of the signaling apparatus, such as basal levels of Rac activation and the expression of the signaling proteins.

The rapamycin stimulation system described here also allows a more detailed study of the interplay between Rac activation and activity of other signaling species, including those that might be involved in various feedback interactions. This analysis is akin to the more common epistasis assays, but with subtler phenotypes related more closely to gradient sensing responses. In particular, our analysis suggested that PI3K interplay with Rac activation, while consistent with the recently proposed formation of an AND gate in terms of the response[6], acts more specifically by controlling the threshold of cell responsiveness to Rac activity gradients. Whereas PI3K inhibition does not prevent the ability of the cells to undergo directed cell polarization or migration responses, it can strongly reduce the fraction of cells capable of these responses within the same set of experimental conditions.

The analysis here represents a more general framework extensible to other rapamycin-activatable signaling molecules, as well as other cell types and multicellular systems. Furthermore, the effects of gradients of other proteins engineered to be sensitive to small, membrane permeable molecules, such as ATP-analogues[48] and imidazole[49], could also be analyzed to refine our understanding of the mechanisms of cell responses to graded intracellular signaling activity. As also demonstrated in this report, such efforts could help develop qualitatively and quantitatively improved mathematical and computational models of gradient sensing and chemotaxis phenomena, extending common approaches to these processes. We suggest that, as the repertoire of methods for direct control of cellular events increases, microfluidics-based tools will play an important role in exploitation of these methods in cell navigation research.

**Acknowledgements**

LEK and WRH were supported by an NSERC discovery and accelerator supplement grants. This work was supported by NIH grants R01GM092930 (T.I., B.L.) and 10052503 (A.L.,B.L.).

**Author Contributions**

BL helped to design the microfluidic device, fabricated devices, ran experiments, analyzed data, and wrote the manuscript. WRH crafted the model in consultation with LEK. TU created the constructs. CW designed the microfluidic device. AH helped in fabricating microfluidic devices. TI concieved the original idea, oversaw various aspects of the project and edited the manuscript. AL participated in all aspects of the project, oversaw its completion and wrote the manuscript.

**Author Information**

Affiliations

Department of Mathematics, University of British Columbia, Vancouver BC, Canada V6T 1Z2

Williams Holmes and Leah Keshet-Edelstein.

Department of Biomedical Engineering, Institue for Cell Engineering, Johns Hopkins University, Baltimore MD, 21218 USA

Benjamin Lin, Chiaochun Wang, Andrew Harwell, and Andre Levchenko.

Department of Cell Biology, Center for Cell Dynamics, Johns Hopkins Medicine, Baltimore MD, 21205 USA

Tasuku Ueno and Takanari Inoue.

**Competing financial interests**

The authors declare no competing financial interests.

**Corresponding Author**

Correspondence should be addressed to: jctinoue@jhmi.edu, alev@jhu.edu

**Figure Legends**

**Figure 1** Graded activation of Rac directs cellular polarity. (**a**) A schematic of the mechanism of Rac activation by rapamycin induced heterodimerization. (**b**) The microfluidic device used to generate linear gradients of rapamycin with a sample image of the microchannels seeded with individual HeLa cells and the corresponding gradient visualized with Alexa 594 dye. The ports are labeled according to function. The red layer of the device is the fluid flow layer, while the green layer is the control valve layer. Alexa 594 dye is used to visualize the gradient (red) in all subsequent images. A sample image of a cell transfected with the Rac activator, YF-TIAM1 experiencing a gradient of rapamycin is shown in the blue box. (**c-f**) Four polarity states observed after the attachment period with respect to the direction of the imposed rapamycin gradient and associated polarization responses to the gradient of rapamycin. Images are rotated by 90° to aid in visualization. Cartoons illustrate the polarity of the associated state and the direction of the gradient. The green color indicates expression of YF-TIAM1. Yellow arrows denote the initial direction of polarity. Times are in minutes. Scale bars, 10 μm. (**g-j**) Kymographs illustrating the morphological changes of the corresponding cells in **c-f** over the experimental period. Blue lines trace the initially polarized face while red lines trace the opposite face. Yellow arrows denote the initial response time and their location indicate which cell face

was the first to change in the gradient. White arrows indicate the late polarization time. Times are in minutes. (**k**) Rac activity visualized by FRET (Raichu-Rac) in a state 3 cell undergoing repolarization towards a gradient of rapamycin. (**l**) Visualization of the Rac activator, MF-TIAM1 in the cell from **k** pre and post rapamycin addition. Times are in minutes. Scale bars, 10 μm. (**m**) Quantification of the difference in FRET response between the front and back of cells over time (n = 9). The front is defined as the side of the cell receiving the higher concentration of the gradient.

**Figure 2** Quantification of the initial response time. (**a**) Sample kymograph of the mathematical model simulation under graded Rac stimulation. (**b**) Model response time vs. normalized gradient ($s_1$) for two different values of $s_o$ (input). (**c-e**) Kymographs chronicling typical changes in cell morphology seen during early time periods. The first image before each kymograph depicts the gradient (red) that the cell is experiencing as visualized with Alexa 594 dye. The gradient is not shown in resulting images to promote clarity of morphological changes. The green color indicates expression of YF-TIAM1. Red dotted lines highlight evolving cell boundaries. Yellow arrows indicate initial polarities. Times are in minutes. Scale bars, 10 μm. (**f-h**) The dependence of initial response times on gradient values. States are color coded; state I (green) n = 29, state II (blue) n = 37, and state III (red) n = 27. In each plot, the colored dots highlight the dependence of that particular state while the grey dots illustrate where the response times of the other states fall. Data is fitted based on simulation results. Spearman correlation coefficient, **f** = −0.679, **g** = -0.655, **h** = -0.583. The asterisk indicates a statistically significant difference (State II vs. State I, p < 1e-4, State II vs. State III, p < 1e-3)

between the state II curve vs. the other states. There is no statistical difference between state I and state III (p = 0.54). Both tests were carried out using an F test.

**Figure 3** Quantification of the late response time. (**a**) Kymograph depicting the change in the lamellipodium directed by the gradient and dimming seen in the cell body during the late polarization time. The image preceding the kymograph illustrates the gradient (red), visualized with Alexa 594, received by the cell. The green color visualizes expression of YF-TIAM1. Times are in minutes. Scale bar, 1 0 μ m. (**b**) Simulations showing response strength vs. input ($s_o$) for two gradient levels ($s_1$); note bifurcations at distinct $s_o$ values. Response strength is defined as ratio of Rac activity at the front vs. the back in the model cell. (**c**) The intensity of the cell body normalized to the initial time point. Intensity values are taken as the mean of the intensity of the area enclosed by the red trace in **a**. The drop line indicates the late polarization time, with a circle highlighting the inversion of response used to define it. (**d**) 3D reconstruction of confocal Z slices of the same cell taken pre- and post- rapamycin addition. The "post-" cell image was taken 240 min after treatment. Scale bar, 10 μ m. (**e**) Cell body volume before and after rapamycin addition. The data shows the mean of n = 9 cells and error bars show SEM. The asterisk denotes a statistically significant difference, p = 0.019, using a two sided student's t-test (**f**) The late response time as a function of mean concentration for different gradient values. State I (green) n = 29, state II (blue) n = 37, and state III (red) n = 27. The response times of other states are superimposed on each plot in grey. The pink drop line in the state III plot demarcates the separation point between unresponsive cells and responsive cells. Pearson correlation coefficient of linear regressions- State I = -0.608, State II = -0.783, State III = -0.698). The red asterisk denotes a statistically significant difference (state II vs. state I, p < 1e-4, state II vs. state III, p <

1e-4) between the y intercept of the linear regression for state II vs. the y intercepts of the regression data from other states. There is no significant difference between the y intercept of state I vs. state III (p = 0.13). Both statistical tests were carried out using an ANCOVA test.

**Figure 4.** A Rac activity threshold underlies the late polarization time. (**a,d**) Time series of representative cells stimulated with a gradient of rapamycin for 30 minutes and 60 minutes respectively. Yellow arrows indicate the initial direction of cell polarity. Times are in minutes. Scale bars, 10 μm. (**b,e**) Kymograph illustrating morphological changes seen in **a**,**c** respectively. Blue lines track the initially polarized cell face while red lines track the opposite face. Below the kymograph is the experimental scheme used to stimulate the cell. (**c,f**) Late polarization time for a 30 minute and 60 minute stimulation period for all three states, respectively. For 30 minutes- State I (green) n = 17, state II (blue) n = 13, and state III (red) n = 31. For 60 minutes- State I (green) n = 32, state II (blue) n = 37, and state III (red) n = 32. Diamonds indicate cells that did not respond within the experimental time frame (240 minutes), while circles indicate responsive cells. Grey dots in each plot represent the late polarization times for cells in each state from Fig. 3. The drop lines demarcate a threshold between responding and non-responding cells.

**Figure 5.** $PIP_3$ feedback modulates Rac mediated polarization. (**a**) Late polarization time dependence against mean concentration with LY294002 treatment. State I (green) n = 21, state II (blue) n = 31, and state III (red) n = 27. Diamonds represent non-responder cells while circles represent responding cells. Grey dots on each plot illustrate the late polarization times seen in Fig. 3. The grey drop line in the state III cell response plot represents the previous response

threshold between non-responding and responding untreated cells, while the colored line represents the shifted threshold following LY294002 treatment. (**b**) Model schematic as in Model Figure 1 with red 'X' indicating that feedback from $PIP_3$ ($f_1$) is decreased during the subsequent simulations. (**c**) Simulations of response strength vs. signal strength for different feedback levels (In all previous simulations, $f_1 = 1$). (**d**) The response threshold for state III cells with and without LY294002 treatment, respectively. The drop lines indicate the response threshold.

**Methods**

**Modeling**

As described in the Supplementary Model file, we have developed a simple model based primarily on the antagonistic relationship between the two small Rho family GTPases involved in polarized cell migration, Rac and Rho, modulated by Cdc42, based on sequential model selection. The model was implemented as a system of partial differential equations (PDEs) for Rho GTPases in 1 space dimension and explored numerically using Matlab® (MathWorks). The model was later extended to include phosphoinositide (PIP, PIP2, PIP3) metabolism. Signal gradient (represented as graded Rac-GEF activation) and its minimum value (as a proxy for the average value explored in the experimental analysis) were used to probe the response of the model to stimulus. PDEs were solved with a fully implicit diffusion / explicit reaction scheme in Matlab. Runs from multiple settings of signal strength, or other parameter variations were pooled together to generate the responses shown in Figs. 2b, 3b, and 5c (See Supplementary Information for details).

**Device Fabrication**

Microfluidic chips were created using a two layer soft lithography process[50]. PDMS (GE RTV) was used to create molds for the devices as described previously[51]. To increase the height of the flow through channels, the membrane between the control layer and flow layer was removed. Before an experiment, each device was cleaned with 70% ethanol and was allowed to bond to a clean 22x40 mm coverslip (Fisher).

**Cell Culture and Transfection**

HeLa cells were maintained in Dulbecco's modified Eagle's medium with 10% FBS and 1% Penicillin Streptomycin (Gibco). MTLn3 cells were cultured in Alpha minimum essential medium supplemented with 5% FBS and 1% Penicillin Streptomycin (Gibco). Both cell lines were kept in a 37°C and 5% $CO_2$ environment during culture and in experiment. MTLn3 cells were provided by the Segall lab. The constructs, YF-TIAM1 and $Lyn_{11}$-FRB were transfected into cells using Fugene HD (Roche) per manufacturer's recommendations. The Raichu-Rac FRET probe was kindly provided by the Matsuda lab and was also transfected in a similar manner. During FRET experiments, MF-TIAM1 was used in place of YF-TIAM1. MF-Tiam1 has a MCherry fluorophore in place of YFP and was used to avoid spectral overlap with the FRET probe.

**Imaging**

Microfluidic experimental imaging was performed using an inverted Zeiss Axiovert 200M epifluorescence microscope under 37°C and 5% $CO_2$, coupled to a Cascade II:1024 EMCCD camera (Photometrics) using a 40x, 1.3 numerical aperture oil immersion objective (Zeiss). The microscope was driven by Slidebook software (Intelligent Imaging Innovations). Images were taken every 5 minutes in the YFP channel using a 494 nm excitation filter and 530 emission filter set (Semrock) and Alexa 594 dye was imaged using a 572 nm excitation filter and 628 emission filter (Semrock) over a four hour period. A spectral 2d template autofocus algorithm was employed between images to account for any focus fluctuations. To correct for uneven illumination, all images were normalized with the following correction $C = (I-D/F-D)*M$ where C = corrected image, I = initial image, D = darkfield image, F = flatfield image, and M = mean of difference between flatfield and darkfield images. The flatfield and darkfield images were taken as averages of multiple images. FRET images were taken using dual camera capture (Photometrics) with a CFP excitation filter (Semrock), an appropriate dichroic (Semrock), and a YFP/CFP emission filter (Insight) which split YFP emission and CFP emission between the two cameras. Volume analysis was performed on an inverted Zeiss Axiovert 200 spinning disk confocal microscope, coupled to a CCD camera (Hamamatsu) using a 40× objective (Zeiss). The microscope was driven by Metamorph 7.5 imaging software (Molecular Devices). YFP excitation was trigged with an argon laser (CVI-Melles Griot) which was fiber-coupled (OZ optics) to the spinning disk confocal unit (CSU10; Yokogawa) mounted with a YFP dichroic mirror (Semrock) and an appropriate YFP filter (Chroma Technology).

**Image analysis**

All analysis was performed using custom written codes in Matlab 2007b (Mathworks). Cell-based data (Cell length, centroid, etc.) was obtained from the YFP images while gradient based data was obtained from the Alexa 594 dye images. Cells were segmented from the YFP channel based on intensity. The signal to noise ratio was sufficiently high to preclude the use of more sophisticated segmentation techniques. Once cell boundaries were determined from the segmentation, concentration lines spanning the width of the channel were generated from the front and back of the cell to obtain local concentration data. The concentration lines were restricted to the width and length of an individual channel by manually selecting the boundaries of the channel from the Alexa 594 dye image. For any cell that extended out of the channel, the concentration lines were restricted to the respective ends of the channel to avoid spurious measurements associated with the height difference between the arms of the flow channels and the cross channels. Dye intensities were extrapolated to the intensities at the respective ends of the channel to determine concentration.

**Volume analysis**

The cell volume was determined by taking a region of interest (ROI) in the cell body from 3D reconstructions carried out using Matlab. The number of pixels in the ROI were then converted to $\mu$m based on pre-calibration of the slice height.

**Measurement of initial response time and late polarization time**

All measurements were performed using Matlab. To determine the length of a cell's front and back portions, the cell nucleus was tracked by manually fitting an ellipse to the nucleus image and taking the centroid of the fitted ellipse as the position of the nucleus. Nuclei were clearly distinguishable in all YFP images analyzed due to the label exclusion. Cell front and back lengths were calculated by taking the coordinates of the front and rear of cells from segmented images and calculating the distance to the nucleus position (Supplementary Fig. 5). To assay the width of the front and rear of each cell, the distance between the rightmost and leftmost coordinates of the cell front and rear were calculated (Supplementary fig. 5). The initial response time was taken as the time to reach 20% of the total magnitude of the first morphological change towards the gradient (Supplementary Fig. 5). The initial response time was taken at the 50% level for MTLn3 cells due to the faster kinetics associated with their responses. To find the late polarization time, the decrease in fluorescence of the cell body was measured as a function of time. The fluorescence intensity was determined in ROIs, first chosen automatically based on the end coordinates of the cell (front and back) (Supplementary Fig. 9). The ROIs were then further eroded by several pixels to avoid any effects from the cell membrane. The late polarization time was taken as the time to reach 50% of the full fluorescence intensity drop in the ROI from the peak intensity value.

**FRET analysis**

FRET images were analyzed by first subtracting background from each individual CFP and YFP FRET image (CFP excitation, YFP emission). The images were subsequently aligned with a dft registration algorithm[52] and the FRET ratio was calculated by dividing the YFP FRET image by the registered CFP image. The front and back FRET ratios (Fig. 1m) were calculated by taking the mean FRET ratio of the pixels above and below the center of the nucleus. Front and back were defined as the faces of the cell receiving the high and low concentration of the rapamycin gradient.

**Population separation**

Thresholds between non-responding cells and responding cells were determined using quadratic discriminant analysis ('classify' function) in Matlab. The two populations along with their corresponding concentrations were input into the function as training data and a separation point was generated from a given vector of concentrations.

**Statistical analysis**

Statistical analysis was carried out with Sigmaplot software (Systat) and Graphpad Prism (Graphpad). Experimental results were expressed as means with error bars equal to standard error of the mean (SEM). Comparisons between two groups were carried out with a two sided student t-test when assumptions of normality were fulfilled. For comparisons which did not satisfy the assumption of normality, a Mann-Whitney Rank Sum test was used. Comparisons were deemed to be significant if p values were < 0.05. The comparison between the y-axis intercepts of the linear regressions of the relevant data points was carried out using an analysis of covariance test (ANCOVA). First, the difference between slopes was compared; if the difference was insignificant, a comparison between the y-axis intercepts was performed. To compare fitted curves, an F-test was conducted, using the standard procedure.

Andre Levchenko Fig. 1

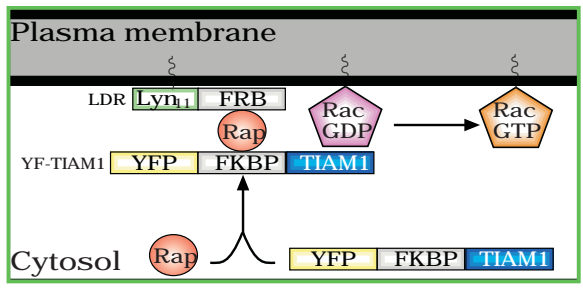
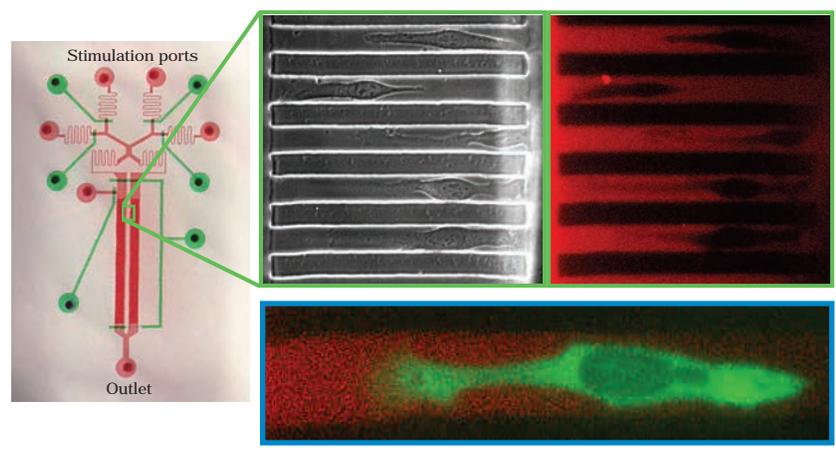
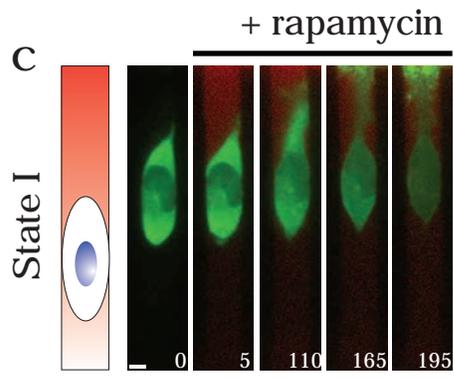
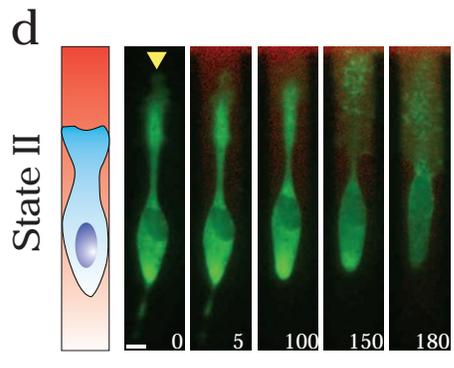
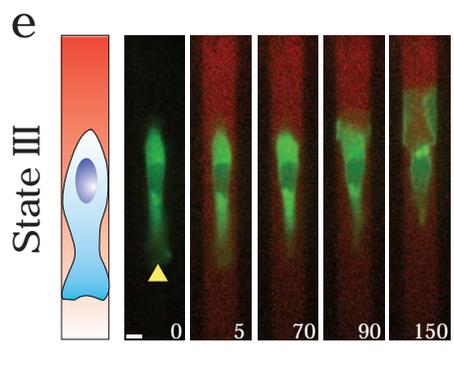
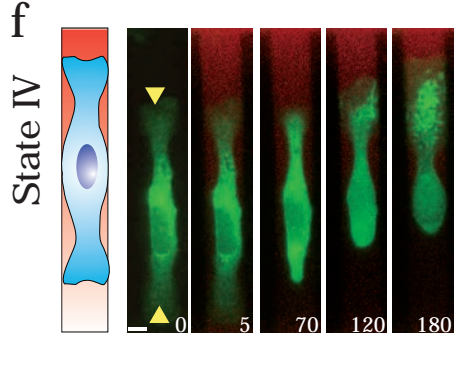
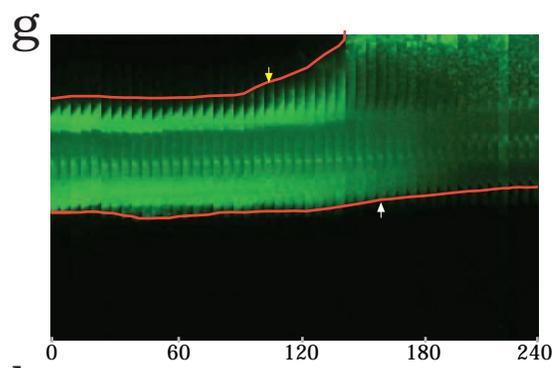
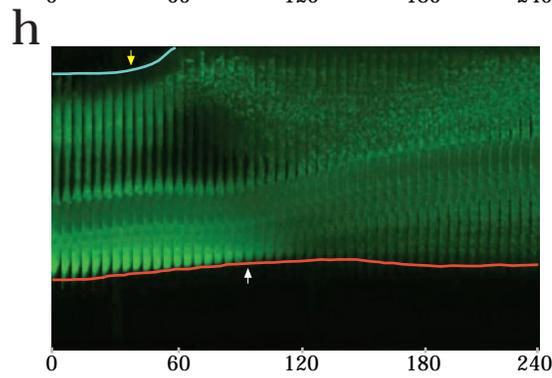
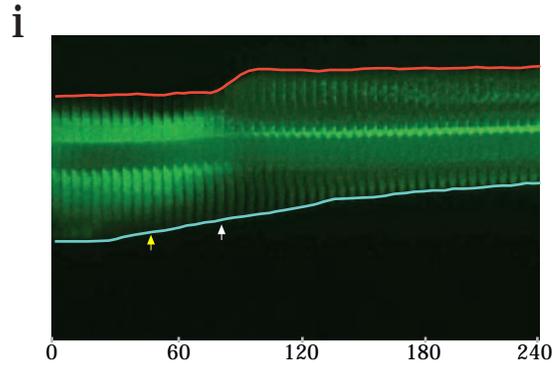
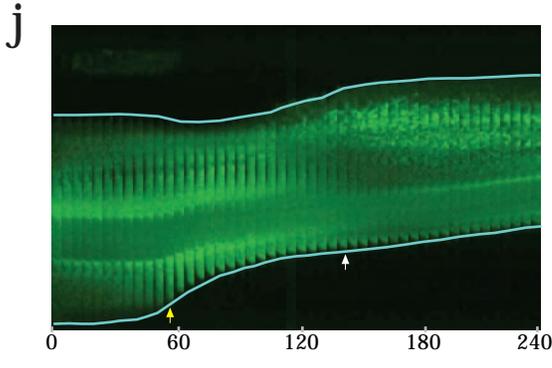
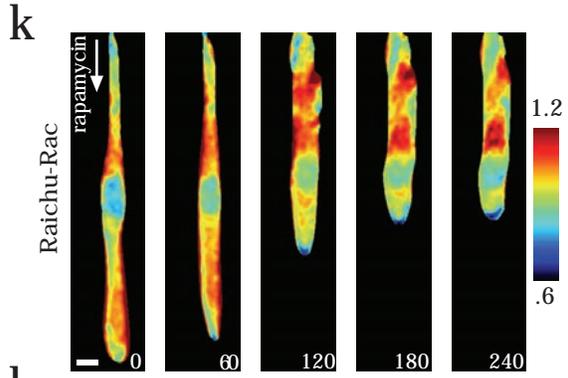
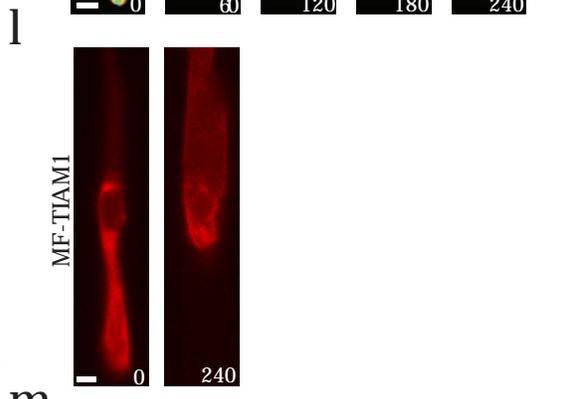
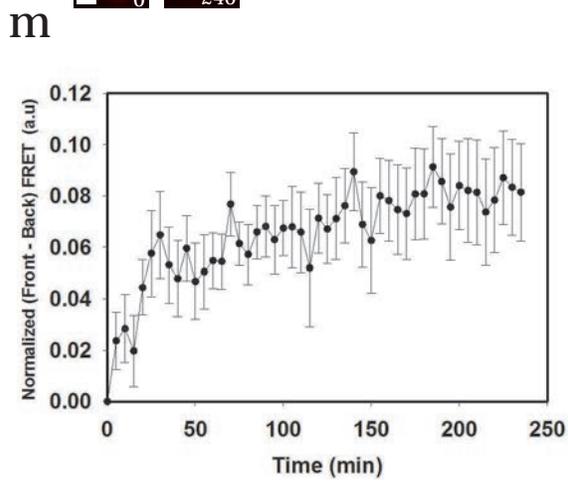

a
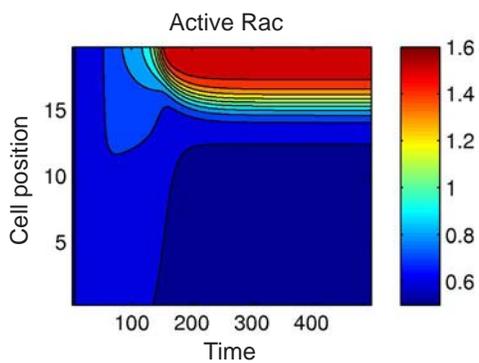

b
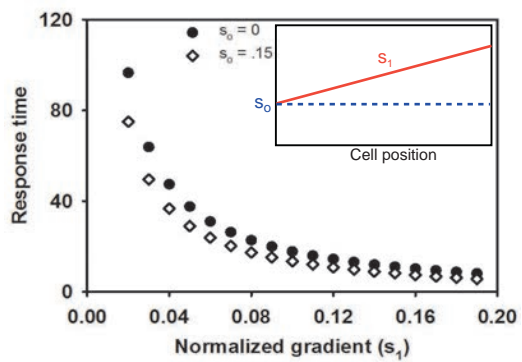

c
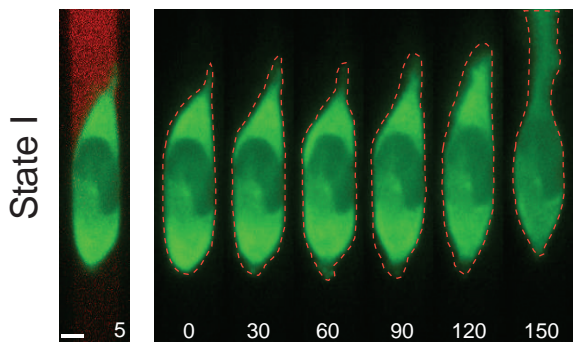

f
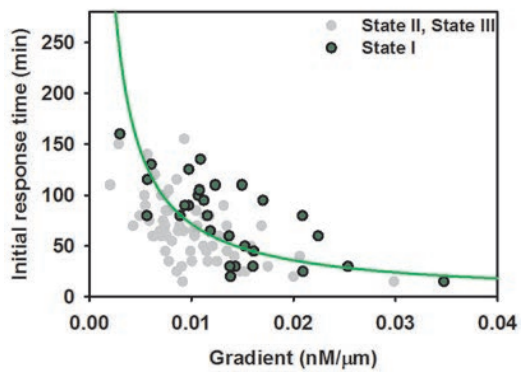

d
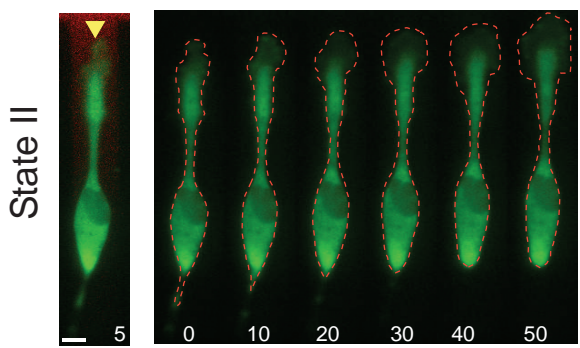

g
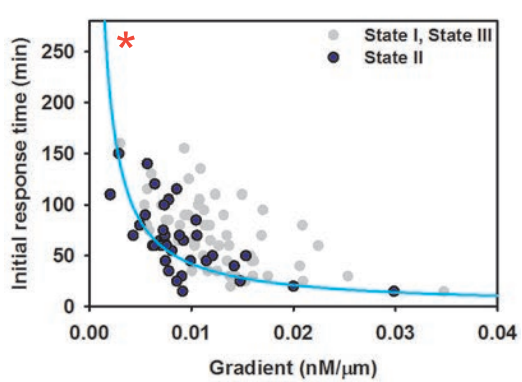

e
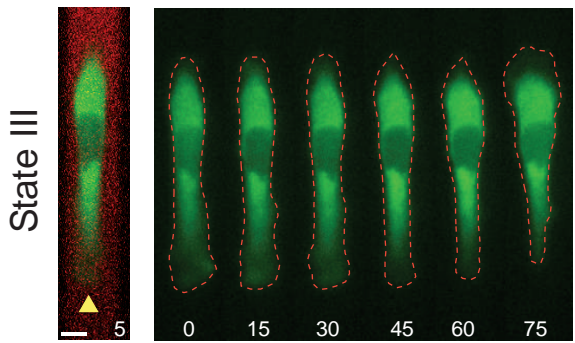

h
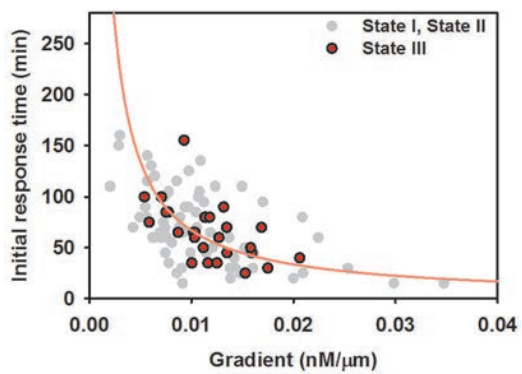

Andre Levchenko Fig. 3

**a**

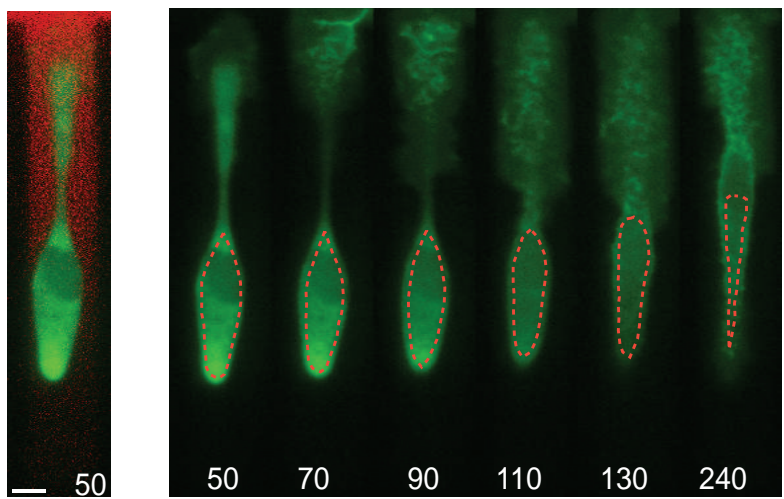

**b**

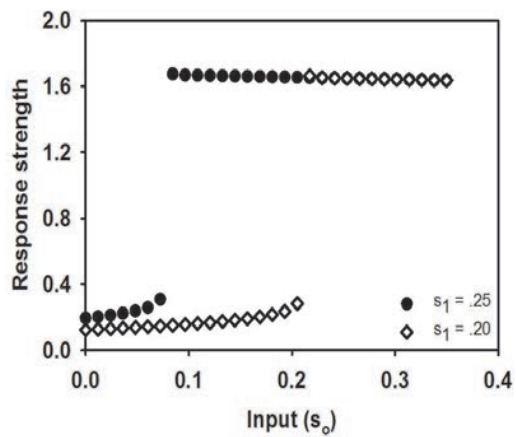

**c**

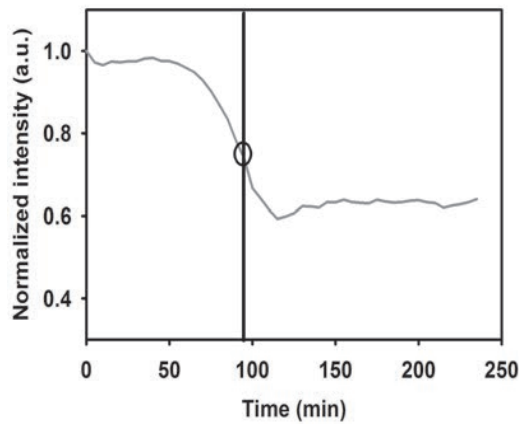

**d**

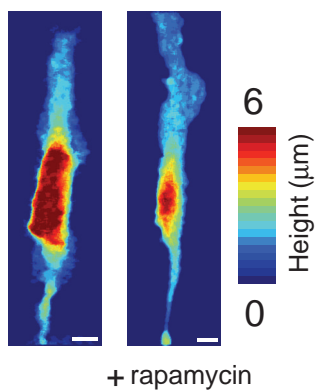

+ rapamycin

**e**

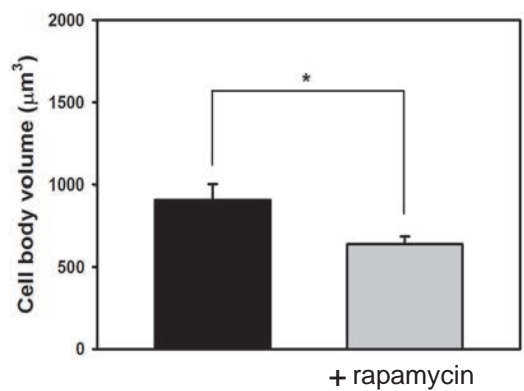

**f**

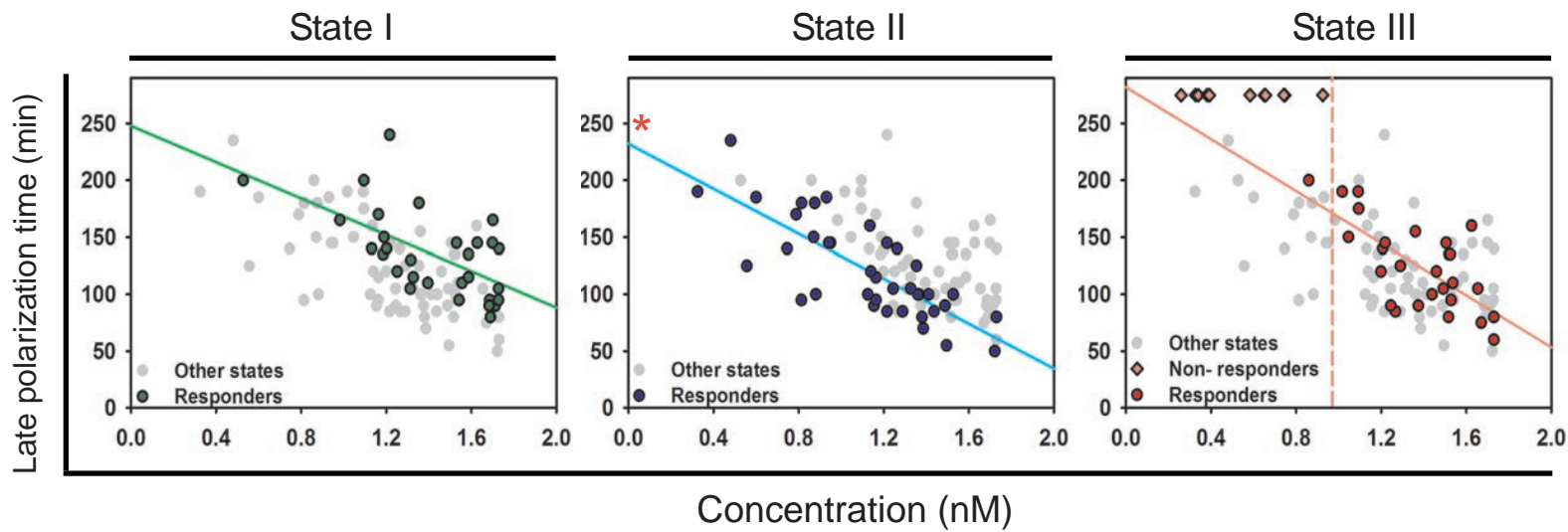

Andre Levchenko Fig. 4

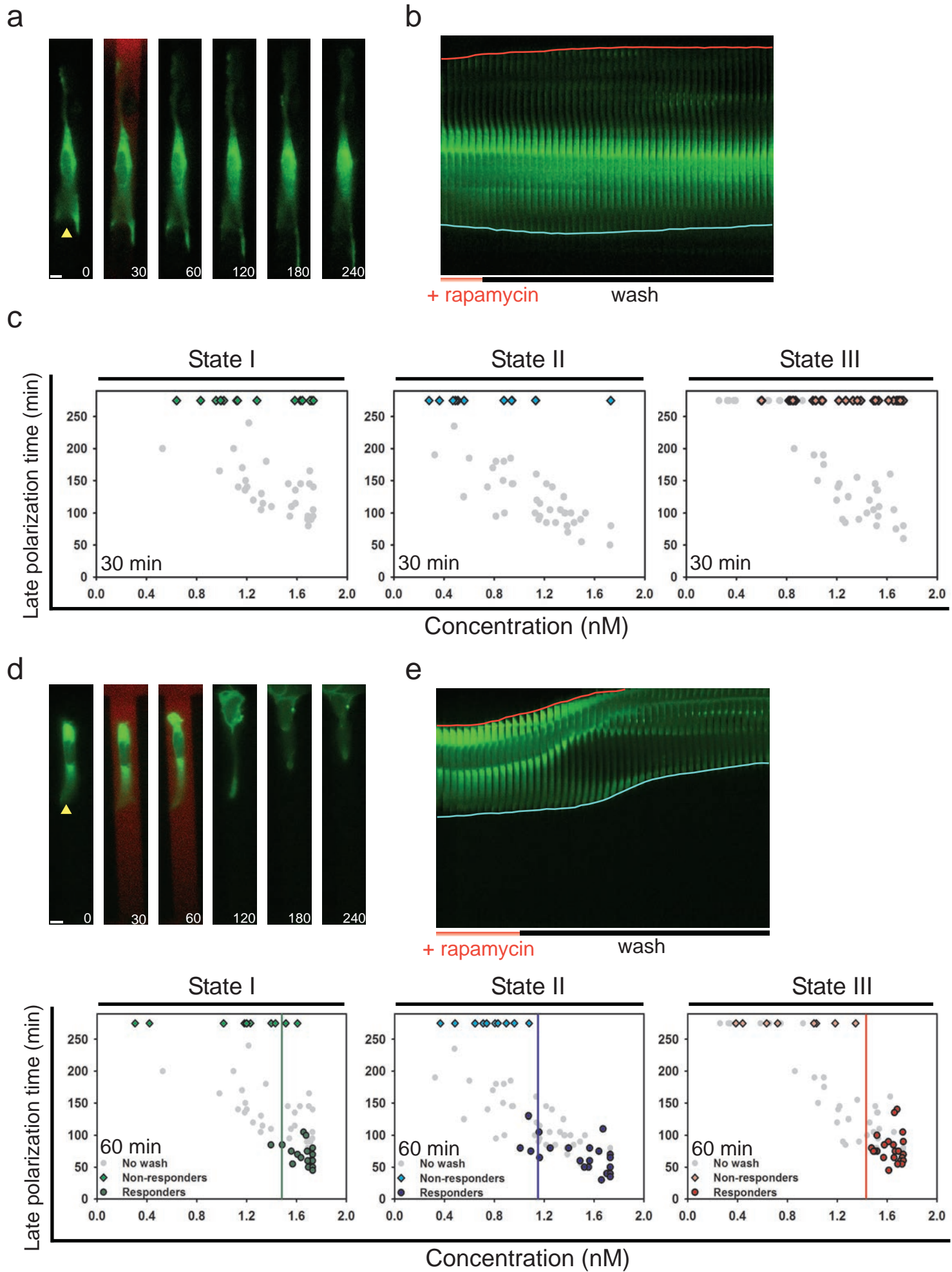

Andre Levchenko Fig. 5

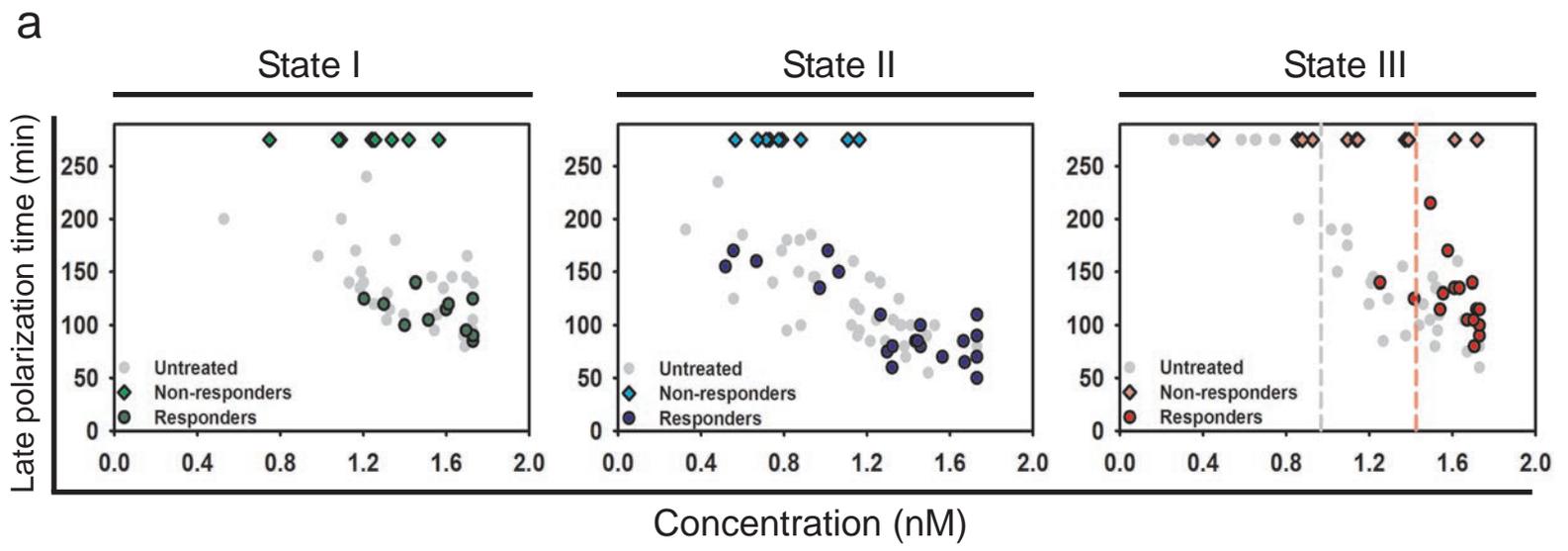

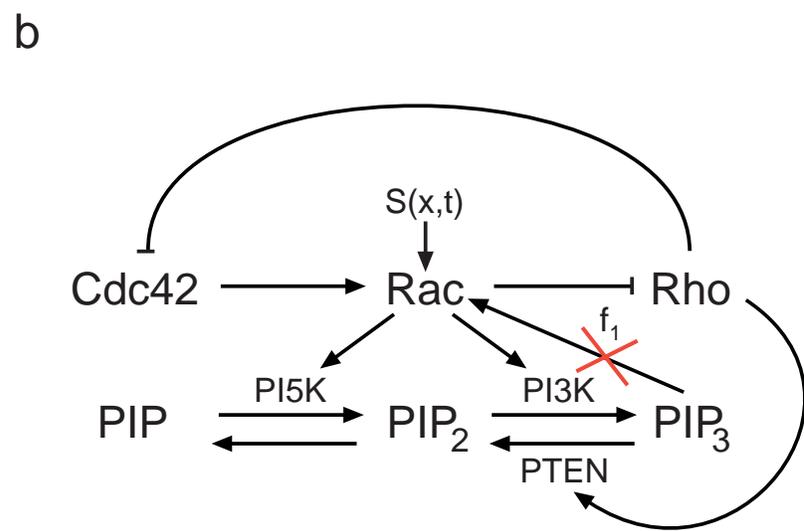

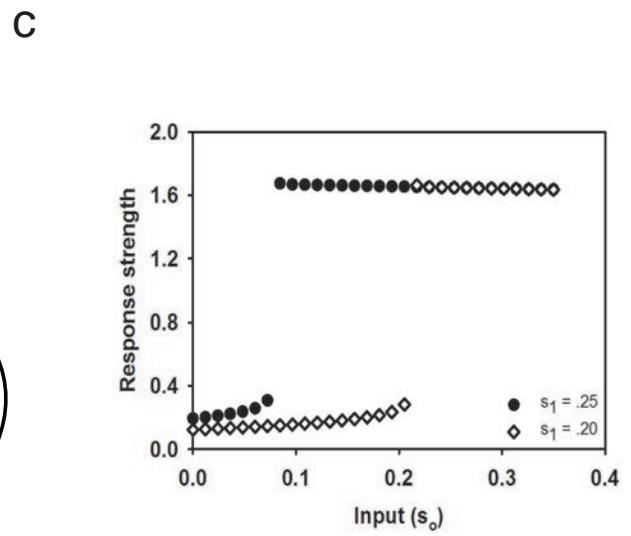

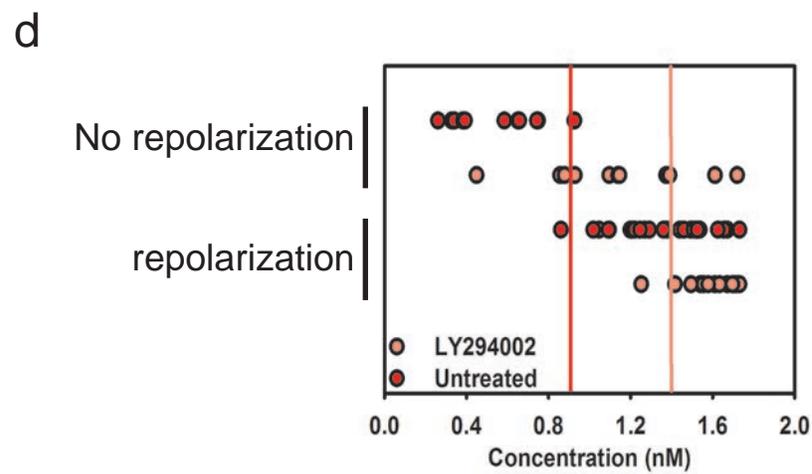

**Synthetic spatially graded Rac activation drives directed cell polarization and locomotion**

Benjamin Lin[1], Williams Holmes[2], ChiaoChun Wang[1], Tasuku Ueno[3], Andrew Harwell[1], Leah Edelstein-Keshet[2], Takanari Inoue[3*], Andre Levchenko[1*]

**Supplementary Notes**

**1. Model Supplement**
**Introduction**

The development of the signaling model shown in Model Figure 1 follows previous work[1,2], with modifications and geometric considerations relevant to the experimental system. More specifically, it was based on a model selection procedure discussed in detail in the manuscript we submitted for reviewers" information and will be submitting for publication in a computational biology oriented publication venue. Here, we provide a brief description of the resulting model and its thorough analysis relevant to the experimental data presented in the main body of the manuscript.

We consider the three GTPase implicated in polarized cell morphology control and chemotaxis, Cdc42, Rac and Rho (as noted in the attached manuscript, in principle the results could also have been accounted for without Cdc42 inclusion into the model, but Cdc42 was included for completeness). For each, we track the levels of active and inactive forms bound to the membrane, $G$, $G^{mi}$, as well as an inactive (GDI- bound) cytosolic form $G^c$, (where $G = C, R, \rho$ respectively). Each variable is a function of time and space in a given cell, whose geometry is explained later. We assume that an inactive GTPase can cycle on and off the membrane between $G^c$ and $G^{mi}$ and that GEF/GAP activity interconverts the membrane bound species $G^{mi}$, $G$. As will be discussed further, we integrate over the depth direction and every point in the remaining domain is considered to have both a cytosolic and membrane components. Crosstalk shown in the figure is directed at GEFs, with enhanced/reduced GEF activity depicted by arrows/inhibitory connections. Linear GAP activity is used in all cases. For phosphoinositides, we track PIP, PIP$_2$, and PIP$_3$ (whose levels are denoted $P_1$, $P_2$, $P_3$), with interconversions (mediated by kinases and phosphatases) that are enhanced by Rac or Rho as indicated in the schematic (Model fig. 1). The feedback from the PIP layer to the GTPase layer is governed by the tunable parameter $f_1$.

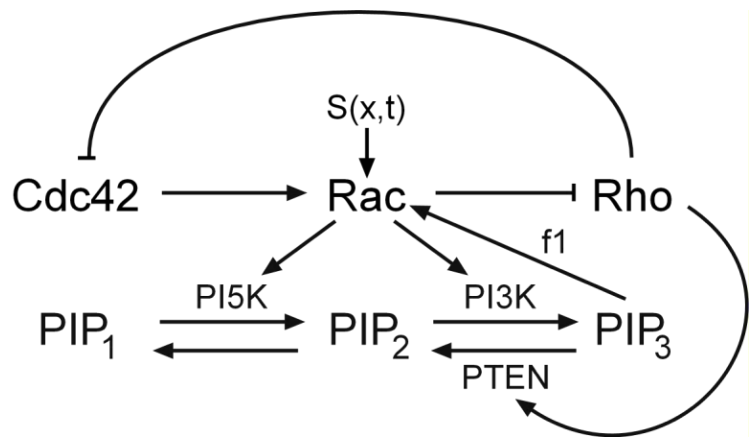

**Model Figure 1**- Diagram of the Rho-GTPase/Phosphoinositide signaling network used in the one dimensional spatial cell model; activation is denoted by (→) and inhibition by (⊥). The signal is $S(x,t) = s_o + s_1 x / L$, where $s_o$ is the input and $s_1$ is the gradient, and L is cell length.

**1.2. Small GTPase module.** Equations for each of the active, and inactive cytosolic and membrane bound GTPases are as follows, for a total of 9 partial differential equations (PDEs).

(1.1-1.3)
$$\frac{\partial G}{\partial t} = I_G - \delta_G G + D_m \Delta G,$$

$$\frac{\partial G^c}{\partial t} = k_{off} G^{mi} - k_{on} G^c + D_c \Delta G^c,$$

$$\frac{\partial G^{mi}}{\partial t} = -I_G + \delta_G G - k_{off} G^{mi} + k_{on} G^c + D_m \Delta G^{mi},$$

where $D_m$, $D_c$ are membrane and cyosolic rates of diffusion, $\delta_G$ is GAP-mediated inactivation rate, $k_{off}$ is the membrane disassociation rate constant, and $k_{on}$ the membrane association rate constant. The term $I_G$ is GEF-mediated rate of activation that will depend on the availability of inactive GTPase, and on crosstalk from other species. Rates of diffusion in membrane and cytosol are estimated as $D_m = 0.1$ μm/s$^2$ and $D_c = 100$ μm/s$^2$ as used previously[2,3]. These equations, supplemented with no-flux boundary conditions conserve the total amount of each GTPase.

**1.3. Geometry and simplification.** Cells are narrowly confined in microfluidic channels, so that their width is constrained and time independent. In view of this fact, it is reasonable to approximate cell shape as a 3D box of length L, width w, and depth d satisfying $d < w \ll L$. Due to controlled signal and the physical constraints of the experimental apparatus, it is reasonable to neglect gradients in all but the length direction. Define a 1D projection of the variable $G^c$ as

(1.4)
$$G^{pc}(x) = \int_0^w \int_0^d G^c(x,y,z)\, dz\, dy \approx dw G^c(x).$$

Here we have approximated $G^c$ as nearly uniform across the width and depth directions. It follows directly that

(1.5)
$$\frac{\partial G^{pc}}{\partial t} = d k_{off} G^{mi} - k_{on} G^{pc} + D_c \Delta G^{pc}.$$

Over the timescale considered, the volume of the cell $V \approx w \cdot d \cdot L$ is roughly constant. Channel diameter determines width w, so the observed lengthening of the cell must be accompanied by depth change. We take the initial values of $d_0 = 0.2$ μm, $L_0 = 20$ μm for a pre-stimulated cell. As L (but not d) is directly observable experimentally, we use $wd = V/L$ to eliminate the less easily measurable cell depth.

**1.4. A composite inactive form.** We consider the cycling of the inactive GTPase between membrane and cytosol to be in quasi steady state, as used previously[1]. We find the fraction of the inactive forms on the membrane and in the cytosol to be be $\gamma(L) := k_{on} / (k_{on} + [V/L] k_{off})$ and $(1-\gamma(L))$, respectively. A composite inactive form, $G^i$, is defined as

$$(1.6) \quad G^i = G^{mi} + G^{pc}, \quad G^{mi} = \gamma(L)G^i, \quad G^{pc} = (1-\gamma(L))G^i.$$

This accounts for the fact that only the membrane bound fraction of this composite is available for GEF activation. We also define an "effective diffusion constant"

$$(1.7) \quad D_{mc}(L) = \gamma(L)D_m + (1-\gamma(L))D_c.$$

This weights the diffusion constant of the composite form according to proportion of time spent on the membrane and the cytosol. A full parameter set is determined by assuming $D_{mc}(L_0) = 50$, consistent with previous work(*1-3*) and $k_{on} = 1s^{-1}$. $k_{off}$ is then determined by

$$(1.8) \quad k_{off} = \frac{L_0 k_{on}}{A} \frac{(D_m - D_{mc}(L_0))}{(D_{mc}(L_0) - D_c)},$$

completing the parameter set associated with membrane cycling.

**1.5. Reduced GTPase model.** The system is now reduced to a set of three GTPases, each described by a single composite inactive form $G^i(x)$ and an active form $G(x)$, the total of which are conserved over the (1D) domain on the experimental timescale. We use the cross-talk depicted in Model Figure 1 to formulate a system of 6 PDEs. Linear inactivation by GAP"s is assumed for each GTPase and up/down regulation of GEF activation pathways are assumed to take generic functional forms leading to

$$(1.10) \quad \begin{aligned} \frac{\partial G}{\partial t} &= I_G \frac{\gamma(L)}{\gamma(L_0)} \frac{G^i}{G_t} - \delta_G G + D_m \Delta G, \\ \frac{\partial G^i}{\partial t} &= -I_G \frac{\gamma(L)}{\gamma(L_0)} \frac{G^i}{G_t} + \delta_G G + D_{mc} \Delta G^i, \end{aligned}$$

with $G = C, R, \rho$ and GEF activation rate functions

$$I_c = \left(\frac{I_c}{1+(\rho/a_1)^n}\right)\frac{\gamma(L)}{\gamma(L_0)}\frac{C_i}{C_t},$$

$$I_c = \left(I_R\left[1+f_1\frac{P_3}{P_{3b}}\right]+\alpha C + S(x,t)\right)\frac{\gamma(L)}{\gamma(L_0)}\frac{R_i}{R_t},$$

(1.11)
$$I_\rho = \left(\frac{I_\rho}{1+(R/a_2)^n}\right)\frac{\gamma(L)}{\gamma(L_0)}\frac{\rho_i}{\rho_t}.$$

A more complete discussion of the forms of these kinetic terms was described earlier[2]. Note that $n \geq 2$ is required for this system to exhibit appropriate polarization behaviour (bistability necessary for a wave pinning polarization to exist[4]). Normalization by $\gamma(L_0)$ is done for convenience of parametrization. Here $f_1$ represents PI feedback strength to the GTPase module via Rac.

**1.6. Phosphoinositide module.** A PI feedback module, based on a modification of earlier work(*1*) by Marée et al (personal communication), is based on the following equations:

$$\frac{\partial P_1}{\partial t} = I_{p1} - \delta_{p1}P_1 + k_{21}P_2 - \frac{k_{PI5K}}{2}\left(1+\frac{R}{R_t}\right)P_1 + D_P P_{1xx},$$

$$\frac{\partial P_2}{\partial t} = -k_{21}P_2 + \frac{k_{PI5K}}{2}\left(1+\frac{R}{R_t}\right)P_1 - \frac{k_{PI3K}}{2}\left(1+\frac{R}{R_t}\right)P_2 + \frac{k_{PTEN}}{2}\left(1+\frac{\rho}{\rho_t}\right)P_3 + D_P P_{2xx},$$

(1.12)
$$\frac{\partial P_3}{\partial t} = \frac{k_{PI3K}}{2}\left(1+\frac{R}{R_t}\right)P_2 - \frac{k_{PTEN}}{2}\left(1+\frac{\rho}{\rho_t}\right)P_3 + D_P P_{3xx}.$$

Terms in round braces are feedback from GTPases Rac and Rho. A full parameter set is outlined in Model Table 1.

**1.7. Simulation Method**. Simulations of the model equations (1.10), (1.11), (1.12) are performed with an implicit diffusion, explicit reaction scheme. Initial GTPase profiles are spatially uniform. The system is allowed to settle to a (parameter dependent) homogeneous steady state by integrating for 50 time units. At $t = 50$, the signal $S(x, t)$ is applied to Rac GEF as shown in Eqn 1.11 and the model is integrated to a final time of 500. We observe that a new polarized steady state emerges on a typical time scale of 100- 300 time units. In all simulations, it was confirmed the rest state was stable with respect to small noise. A kymograph showing the results of a sample simulation is shown in Figure 2a. GTPase asymmetry/polarization strength, shown in Figures 3b and 5c, is measured as the absolute difference between the highest and lowest active Cdc42 (*C*) levels at the final time. (*R* or $\rho$ can also be used to quantify polarization with similar results.) Response times, shown in Figure 2b, are computed assuming a generic inverse relationship, response time = 1/response strength. This represents an assumed inverse relationship between the rate of cytoskeletal reorganization and GTPase polarization strength.

| Parameter Name | Value | Definition |
| --- | --- | --- |
| $C_t, R_t, \rho_t$ | 2.4, 7.5, 3.1 | Total levels of Cdc42, Rac, and Rho |
| $\hat{I}_C, \hat{I}_R, \hat{I}_\rho$ | 2.95, 0.2, 6.6 | Cdc42, Rac, and Rho activation rates |
| $a_1, a_2$ | 1.25, 1.0 | Cdc42 and Rho half-max inhibition levels |
| $n$ | 3 | Hill coefficient for inhibitory connections |
| $\alpha$ | 0.65 | Cdc42-dependent Rac activation |
| $\delta_C, \delta_R, \delta_\rho$ | 1.0 | GAP decay rates of activated Rho-proteins |
| $I_{P1}$ | 10.5 | $PIP_1$ input rate |
| $\delta_{P1}$ | 0.21 | $PIP_1$ decay rate |
| $k_{PI5K}, k_{PI3K}, k_{PTEN}$ | 0.084, 0.00072, 0.432 | Baseline conversion rates |
| $k_{21}$ | 0.021 | Baseline conversion rate |
| $P_{3b}$ | 0.15 | Typical level of $PIP_3$ |
| $L_0$ | 20 | Cell Length |
| $D_m, D_{mc}(L_0), D_P$ | 1.0, 50.0, 5.0 | Diffusion Rates |

**Model Table 1**- Parameter set used for model simulations

## 2. Experimental setup

First, control valves (marked blue in Supplementary Fig. 1a) were primed by connecting syringes filled with deionized water and pressurized to 20 psi. Experimental solutions were made using DMEM F-12 (Gibco) as a base medium. In a typical experiment, two solutions were injected into the device, one containing rapamycin and the other without. Rapamycin concentrations in microfluidic chips were titrated to uniform stimulation experiments in open chambers by comparing temporal responses. The rapamycin solution was injected into inlet "1" (Marked red in Supplementary Fig. 1a). Once the solution reached the intersection of the channel and valve "1", the valve was pressurized to keep the solution pinned at this location. In the corresponding „0" inlet, the rapamycin-free solution was injected and allowed to flush through inlets marked „wash" to wash away any rapamycin which may have entered into the channel. After the wash period, plugs were placed into the "wash" inlets. For experiments utilizing LY294002, a 10 µM concentration of the drug was introduced into both experimental solutions. In all experiments, a 10 µg ml$^{-1}$ fibronectin solution was injected into the „cell" inlet prior to cell introduction. Actuation of valve marked "5" forced the fibronectin solution through the cross channels into the right flow through chamber and out of the outlet labeled „out". Fibronectin coating was performed for 50 min at 37°C. After the coating period, another rapamycin-free solution was placed into the outlet and the fibronectin solution was removed from the „cell" inlet. A HeLa cell suspension, at a concentration of 1 x 10$^6$ cells ml$^{-1}$, was injected into the „cell" inlet with a loading pipette. The outlet pressure was lowered below the atmospheric one, causing flow of the cell suspension and subsequent seeding of cells into the cross channels via the same principle used above for the fibronectin coating. Excess cells in the main flow arms were washed out. In experiments utilizing LY294002, a rapamycin and LY294002 free solution was injected into one of the „wash" inlets to flush away excess cells to prevent prior exposure of cells to the inhibitor. HeLa cells were allowed to attach for 4 h at 37°C and 5% $CO_2$.

**Supplementary Movie 1**

This movie demonstrates the polarization responses of cells across different polarity states to gradients of rapamycin. The green color of each cell represents expression of YF-TIAM1, while the red color illustrates the gradient of rapamycin. Each cell polarizes towards the high side of the gradient. (QuickTime; 5.6MB).

**Supplementary Movie 2**

This movie illustrates the redistribution of active Rac for a representative cell repolarizing in response to a gradient of rapamycin. The pseudocolor depicts the FRET ratio values with higher values having warmer colors and lower values having cooler colors. The cell that is initially polarized in the direction opposite to the gradient repolarizes towards the gradient and active Rac accumulates towards the newly formed lamellipodium and depletes from the previous lamellipodium. (QuickTime; .5MB).

**Supplementary Movie 3**

This movie depicts a cell experiencing a gradient of rapamycin followed by a uniform stimulation of rapamycin. The green color of the cell represents expression of YF-TIAM1, while the red color illustrates the gradient of rapamycin. The cell is visualized for 60 minutes pre-stimulation, followed by a gradient for 120 minutes, and a uniform stimulation for 180 minutes thereafter. During the gradient stimulation, the cell shows enhancement of the cell front and motility towards the gradient. After the switch to a uniform stimulation, the cell begins spreading at the cell rear and loses the enhancement at the front. (QuickTime; 2.1MB ).

**Supplementary Movie 4**

This movie illustrates a cell experiencing a gradient of rapamycin for a period of 30 minutes followed by a washout for the remaining time period. The green color of the cell represents expression of YF-TIAM1, while the red color illustrates the gradient of rapamycin. As seen here, the cell shows little morphological change and does not respond to the gradient within the time frame. (QuickTime; 1.1MB ).

**Supplementary Movie 5**

This movie shows cells in different polarity states experiencing a gradient of rapamycin for a period of 60 minutes followed by a washout for the remaining time period. The green color of each cell represents expression of YF-TIAM1, while the red color illustrates the gradient of rapamycin. Each cell here shows polarization towards the gradient with morphologies similar to those seen in Supplementary Movie 1. Responses continue even when the rapamycin is washed out. (QuickTime; 1.8MB).

**Supplementary Movie 6**

This movie depicts cells in different polarity states experiencing a gradient of rapamycin with the addition of LY294002, a PI3K inhibitor. The green color of each cell represents expression of YF-TIAM1, while the red color illustrates the gradient of rapamycin. Cells did not polarize towards the gradient of rapamycin and remained quiescent throughout the experimental period. (QuickTime; 2.1MB).

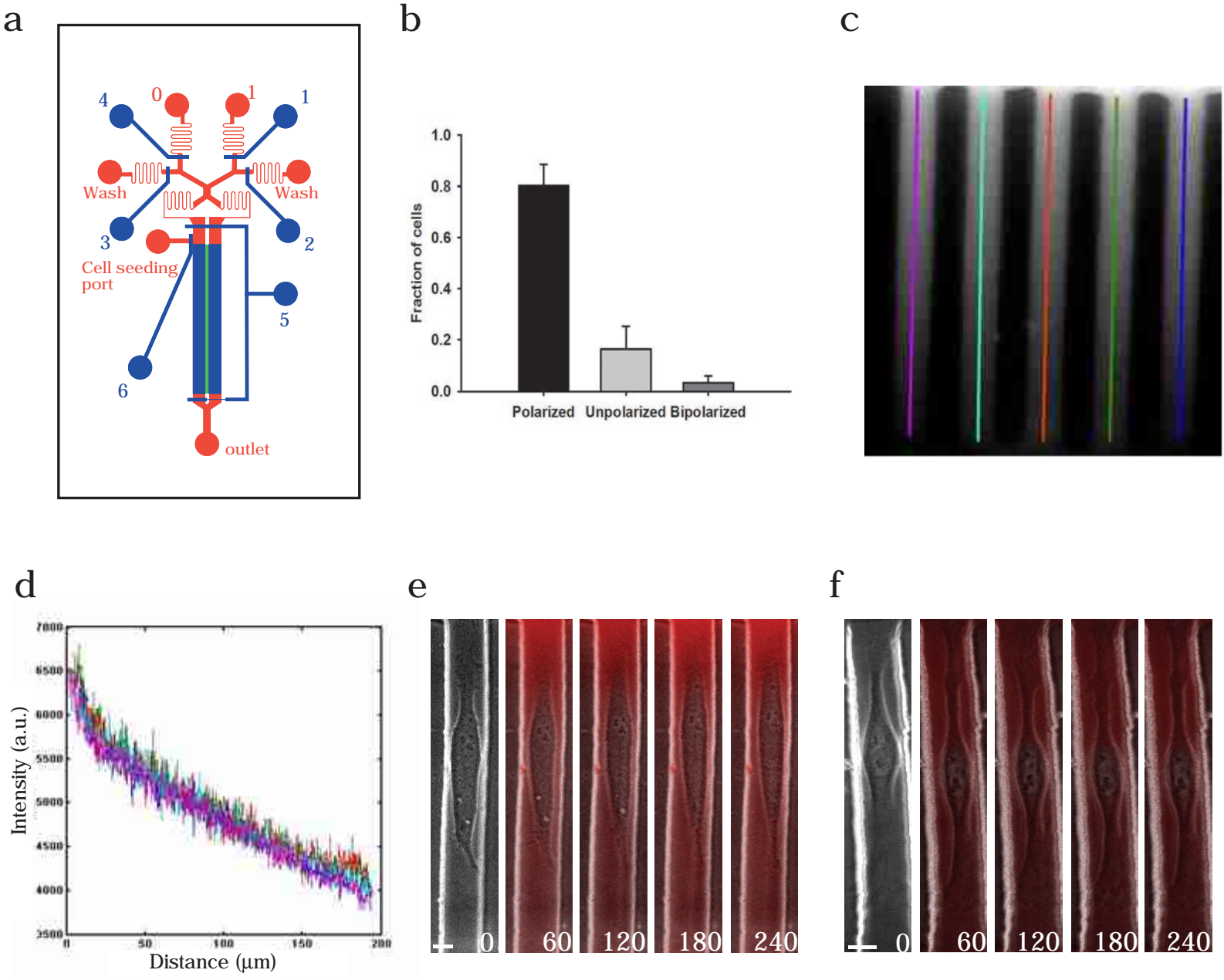

Supplementary figure 1 | Characterization of the microfluidic device. (a) The schematic of the design of the microfluidic device used in experiments.The fluidic layer labeled in blue is the control layer containing fluidically actuated valves used to control the adjacent but separate fluidic layer, labeled in red. The green layer indicates the region where the microchannels are housed and where cells are seeded. (b) Quantification of the fractions of HeLa cell polarity phenotypes observed after a 4 h of cell introduction and attachment period. Data are presented as the mean from n = 8 experiments, with error bars representing the standard error of the mean (SEM). (c) Illustration of Alexa 594 dye in the microchannels with colored lines indicating areas of fluorescence intensity quantifications in (d) which demonstrate the linearity of the gradient profiles and consistency of the gradients in multiple adjacent channels. (e-f) Representative examples of untransfected HeLa cells experiencing a gradient of the dye as in (c), with added rapamycin (e) and without added rapamycin (f). The cells show limited morphological changes through a four hour period. Time values are in minutes. Scale bars, 10 μm.

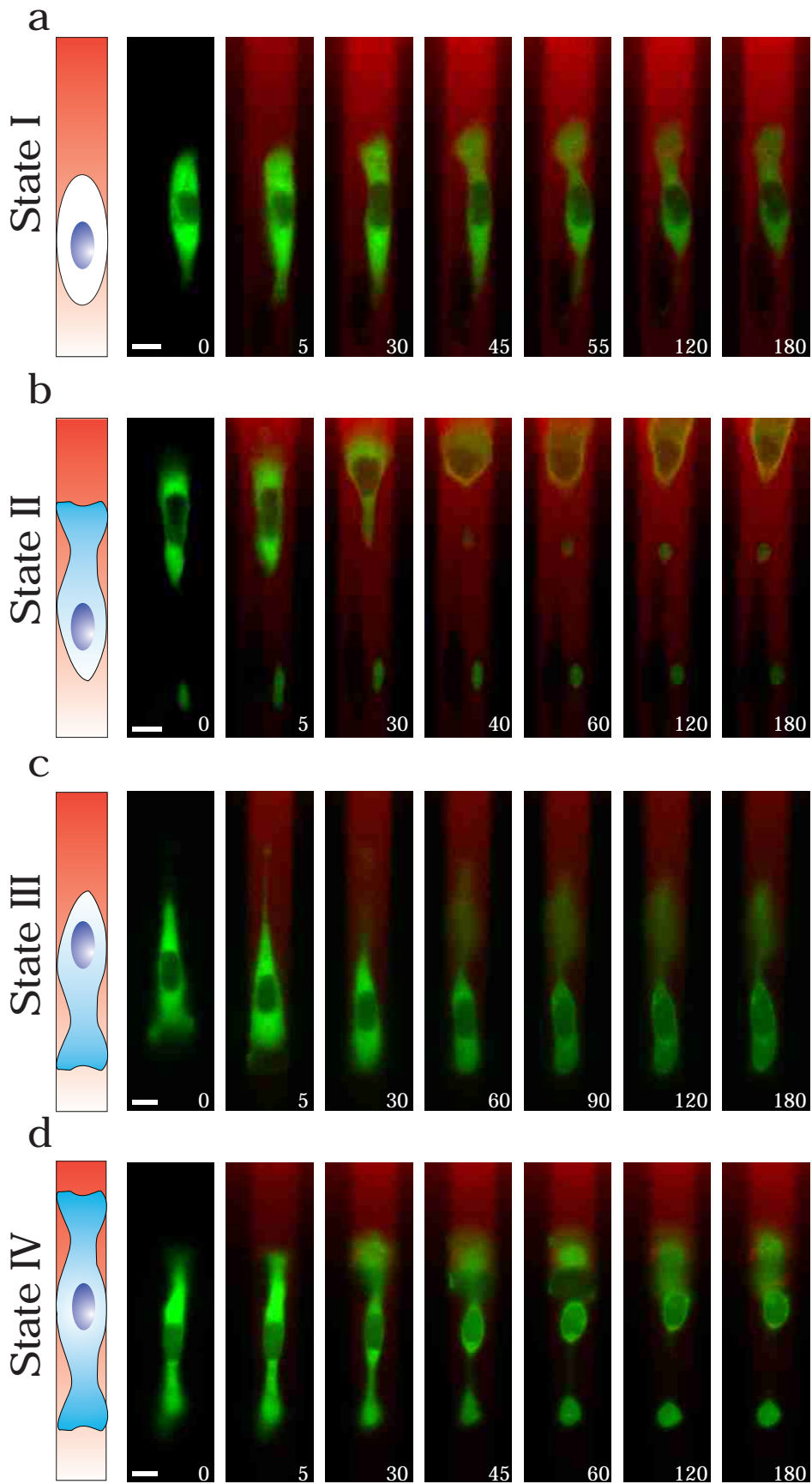

Supplementary figure 2 | MTLn3 cell responses to Rac gradients (a-d) MTLn3 cells in various initial polarity states, similar to those seen in HeLa cells, become polarized towards gradients of rapamycin. The green color indicates expression of YF-TIAM1, while the red color indicates the rapamycin gradient. Times are in minutes. Scale bars, 10 μm.

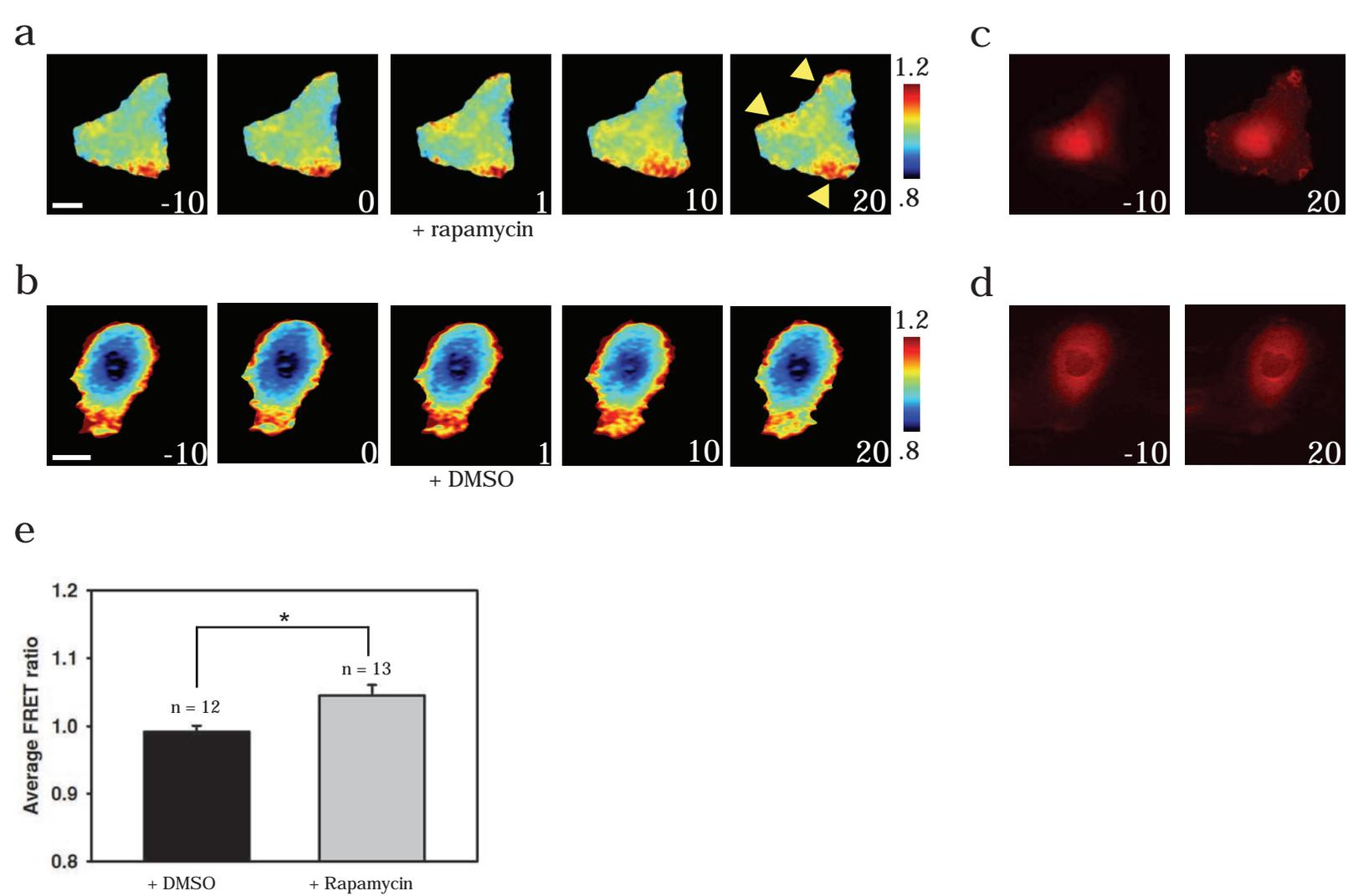

Supplementary Figure 3 | Verification of the Raichu-Rac probe (a-b) Images of the FRET ratio of cells seeded on coverslips following rapamycin addition (a) and rapamycin-deficient solution (b) addition. Scale bars, 20 μm. Yellow arrowheads indicate areas of increased FRET activity. (c-d) Verification of expression of MF-TIAM1 and its membrane translocation post rapamycin treatment (c) and lack of change following rapamycin-deficient solution treatment (d). This construct is similar to YF-TIAM1 construct used in most of the experiments, but has the mCherry fluorescent tag used to avoid spectral overlap with the Raichu-Rac FRET probe. These images verify the efficacy of rapamycin addition. (e) Whole cell average FRET ratio taken 20 minutes after rapamycin or control solution addition. Numbers indicate total cell number for each treatment. Error bars are standard errors of the mean (SEM). The asterisk indicates that the two treatments yielded statistically different results according to Mann-Whitney Rank Sum test ($p < 0.01$).

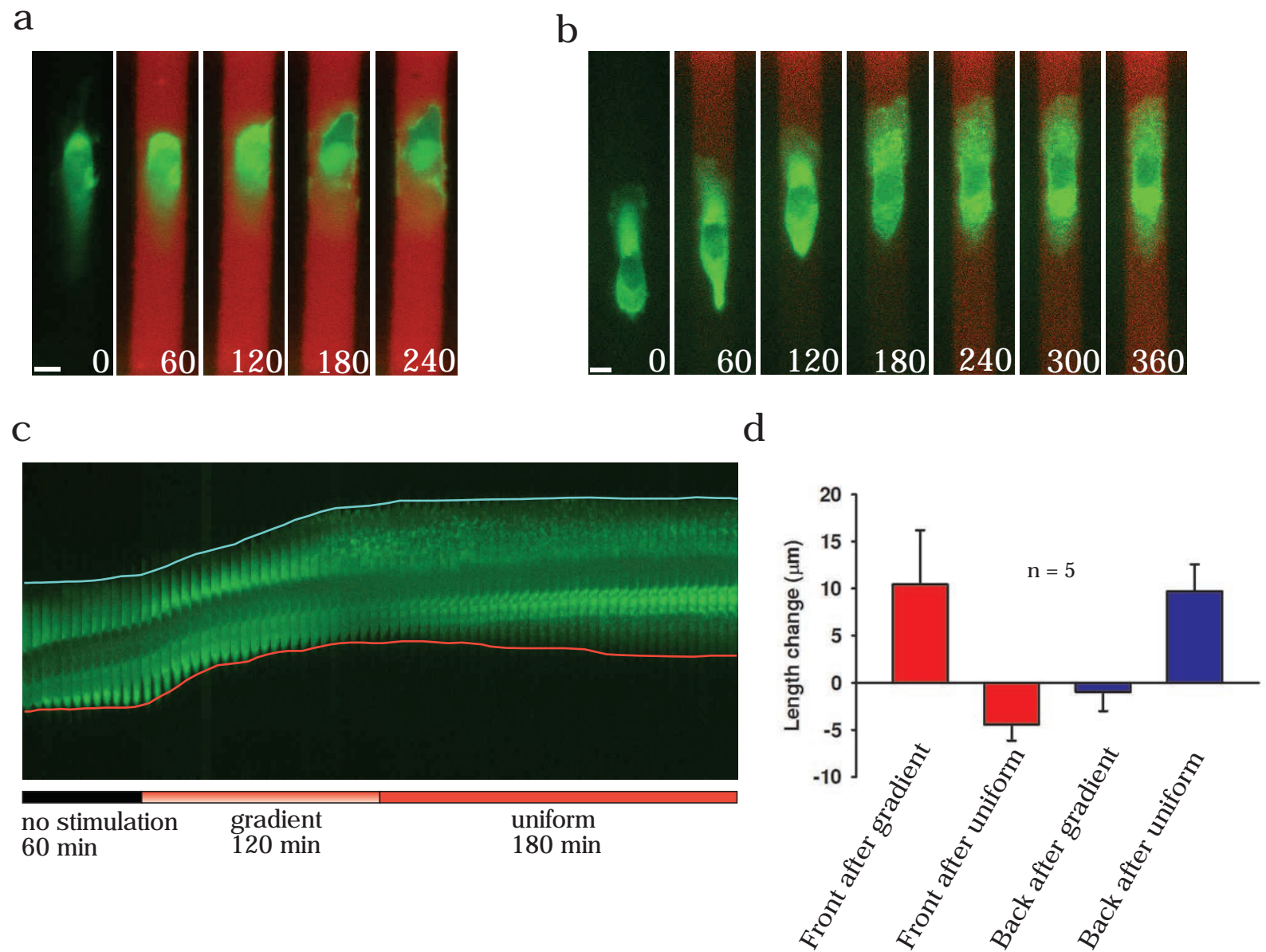

Supplementary Figure 4| Uniform rapamycin stimulation leads to cell flattening and loss of polarity. (a) Cells exposed to a uniform concentration of rapamycin (2 nM) flatten and do not show any directed polarity. (b) Cells observed for one hour without stimulation and subsequently exposed to a gradient of rapamycin for a two hour period, followed by a switch to a uniform stimulation for three hours. The gradient directs cell motility and amplifies the existing lamellipodium. The switch to a uniform stimulation leads to a protrusion formation in the rear of the cell and retraction of the previously amplified lamellipodium. Time values are in minutes. Scale bars, 10 μm. (c) A kymograph depicting the morphological changes of the cell shown in panel (b) over various treatment periods. The blue line traces the front of the polarized cell, whereas the red line traces the unpolarized face. (d) Quantification of the length changes in cell front and back during the rapamycin gradient exposure and after the switch to uniform rapamycin stimulation. Numbers indicate total number of cells. Error bars are SEM.

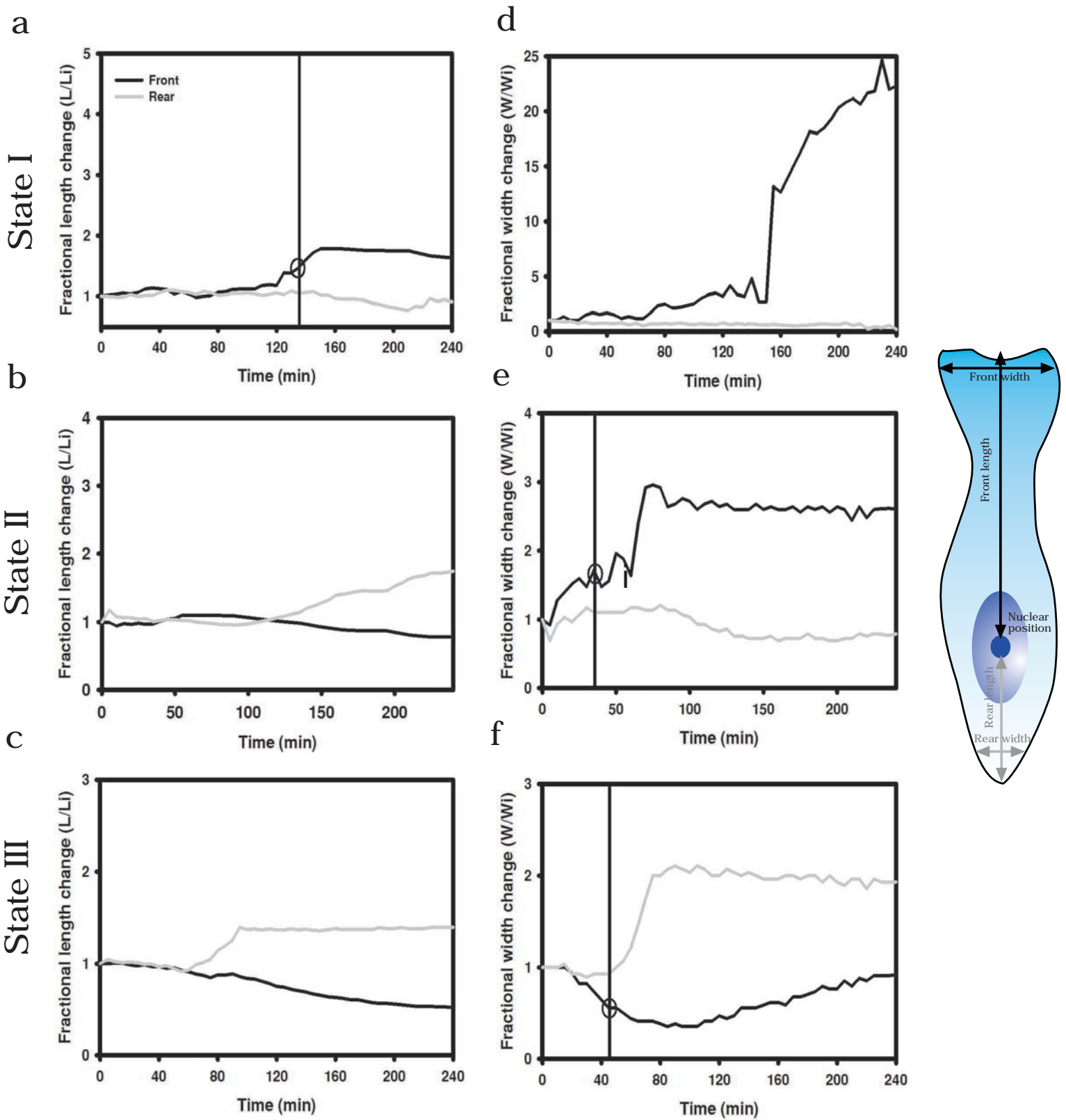

Supplementary Figure 5 | Tracking changes in cell morphology to assay the initial response time. The plots correspond to the cells in different states, as seen in Figure 2. Various morphology metrics are illustrated in the schematic accompanying the graphs. Fractional values of the metrics are shown, with values normalized to those at the beginning of the analysis. (a-c) Tracking of changes in the lengths of the cell front and back over time during gradient stimulation. (d-f) Tracking of changes in widths of the cell front and back. The time to reach 20% of the maximum magnitude of the first morphological response to gradient exposure was taken as the initial response time. The drop line indicates the initial response time, while the circle indicates the intercept between the dropline and the corresponding metric value.

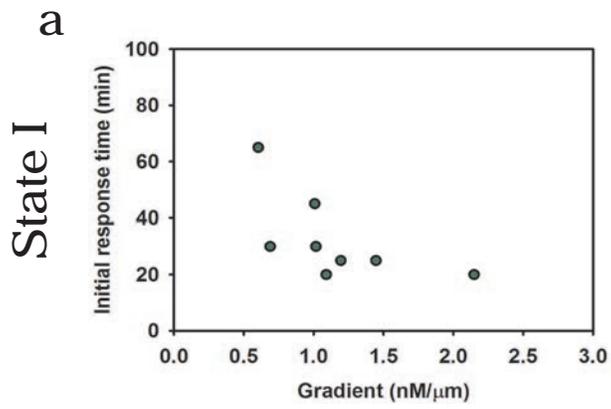
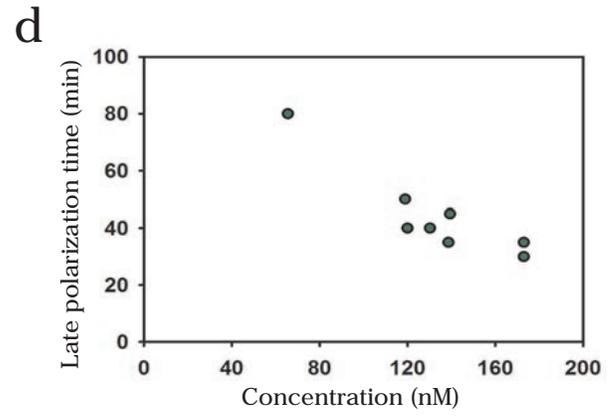
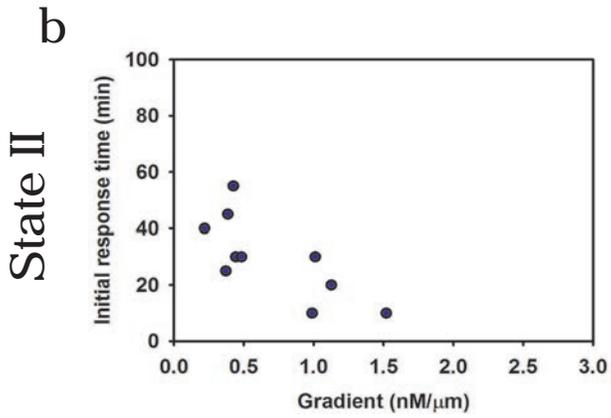
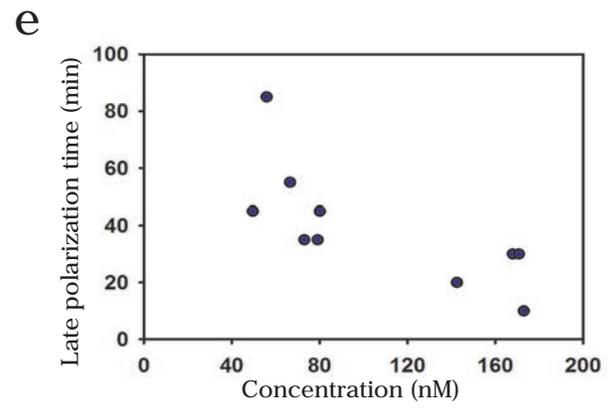
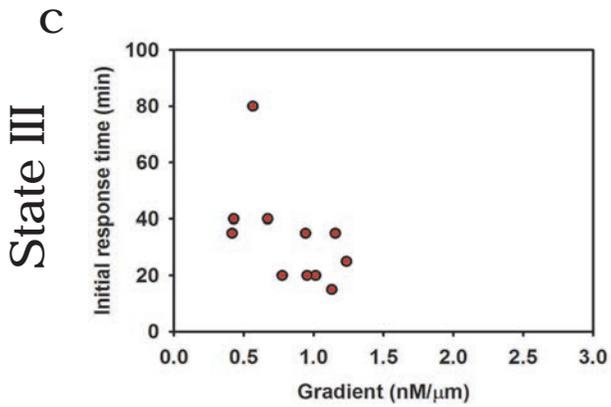
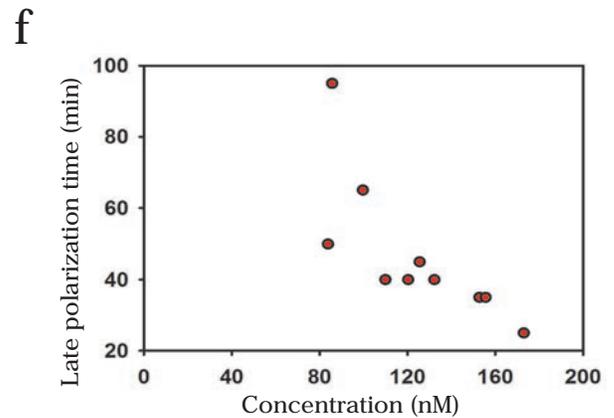

Supplementary figure 6 | Initial response and late polarization times of MTLn3 cells. (a-c) Initial response times as a function of gradient values for cells in state 1, 2, and 3 respectively. (d-f) Late response times for cells in state 1, 2 and 3, respectively as a function of mean concentration. State 1 (a,d) n = 8, State 2 (b,e) n = 10, State 3 (c,f) n = 11.

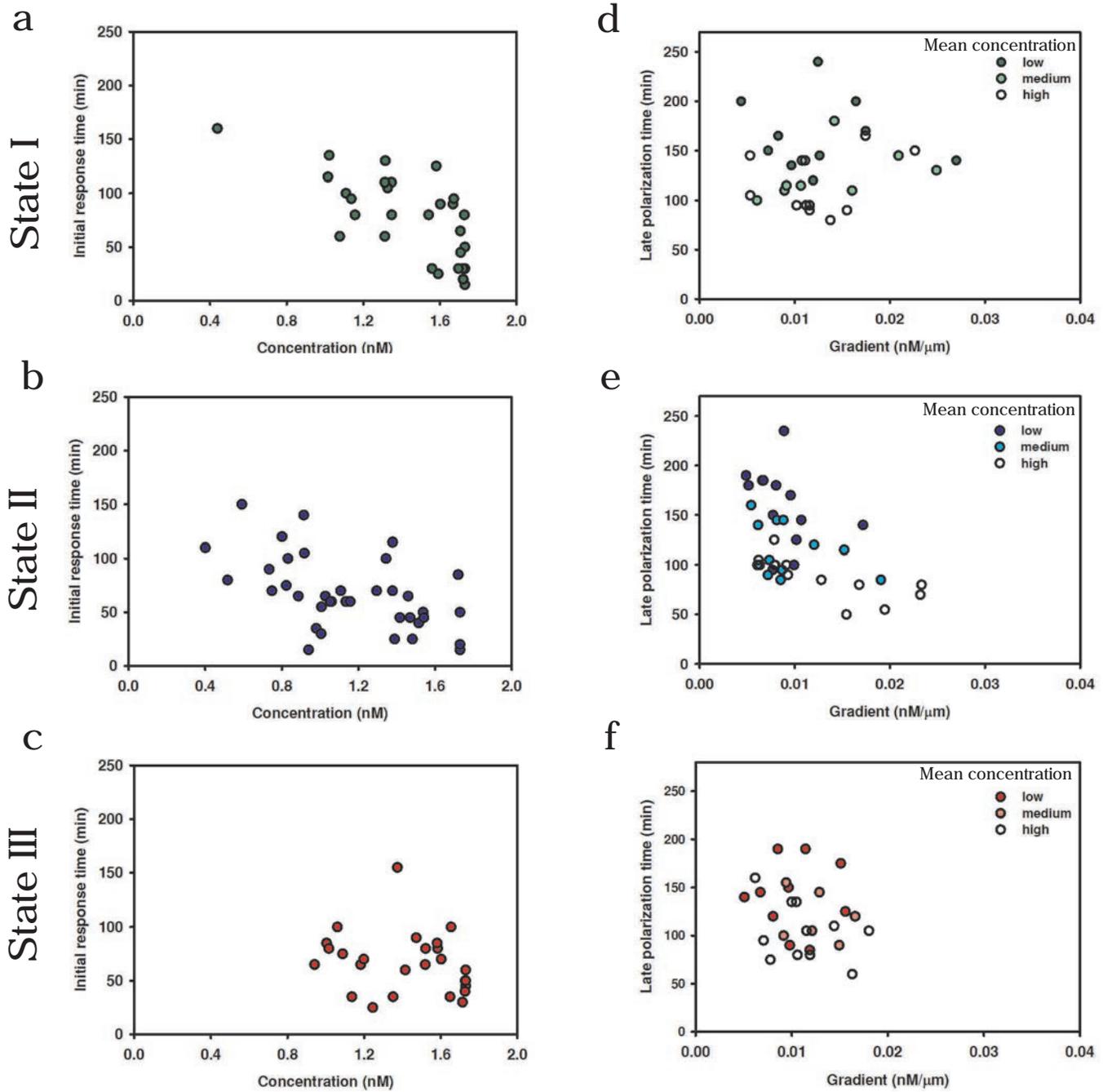

Supplementary Figure 7 | Additional information on dependence of the initial response time and late polarization time on rapamycin gradient and local concentration values. (a-c) The relationship between initial response time and rapamycin concentration for all states. Spearmann correlation coefficient values for each curve were determined as follows, state I: -0.582 (a), state II: -0.478 (b), state III: -0.377 (c). (d-f) Late polarization times determined for different gradient values for cells in all states. Pearson correlation coefficient values for each curve were determined as follows , state I: 0.103 (d), state II: -0.511 (e), state III: -0.228 (f). Within each plot, the data is binned into three mean concentration levels. State I (green) n = 27, state II (blue) n = 37, and state III (red) n = 29.

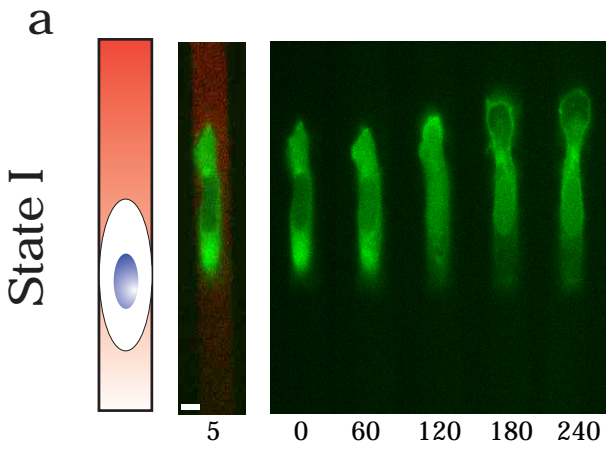
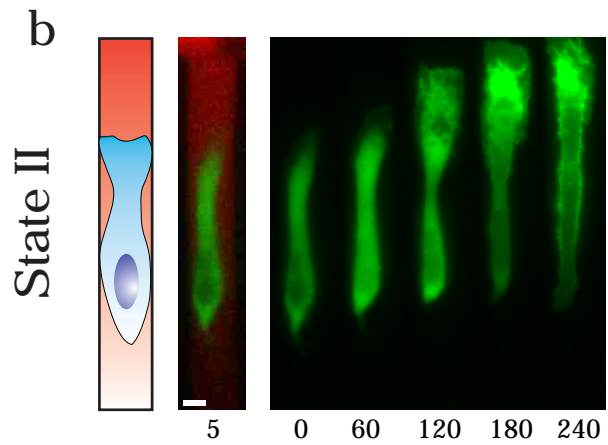
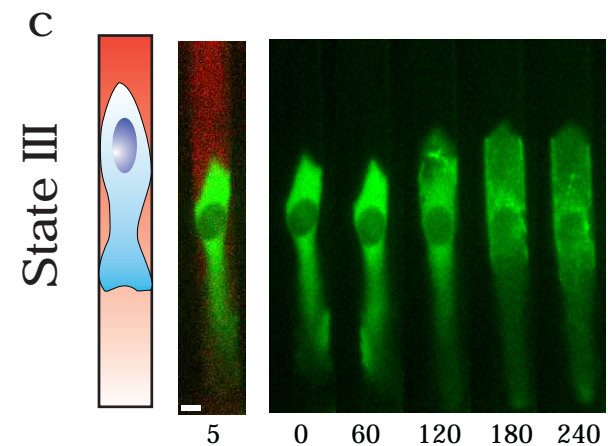
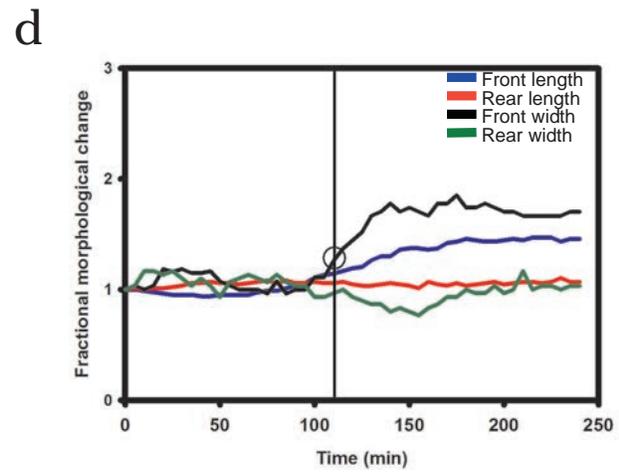
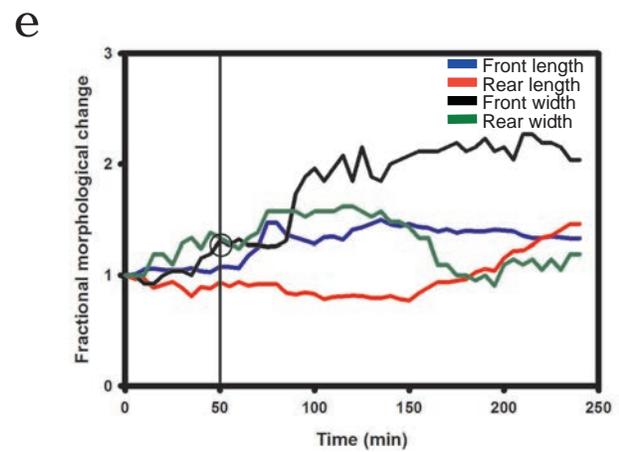
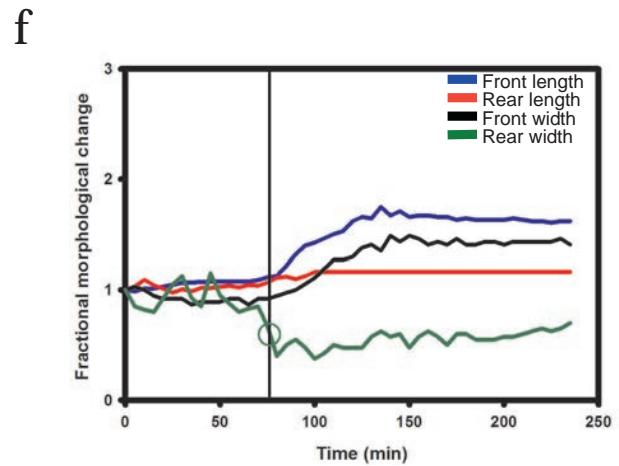

Supplementary Figure 8 | State II cells undergo the initial response faster than cells in other states. (a-c) Examples of the morphological changes in cells in different states experiencing the same gradient of rapamycin (0.01 nM/μm +/- 0.001). Cells were exposed to the gradient for the entire experimental period but only the 5 minute image showing both cells and the gradient visualization is shown for each state to add clarity. Times are in minutes. Scale bars, 10 μm. (d-f) Quantification of the morphological changes seen in the examples shown in (a-c). The dropline indicates the initial response time. The circle indicates the metric used to determine it, with the color of the circle corresponding to the metric.

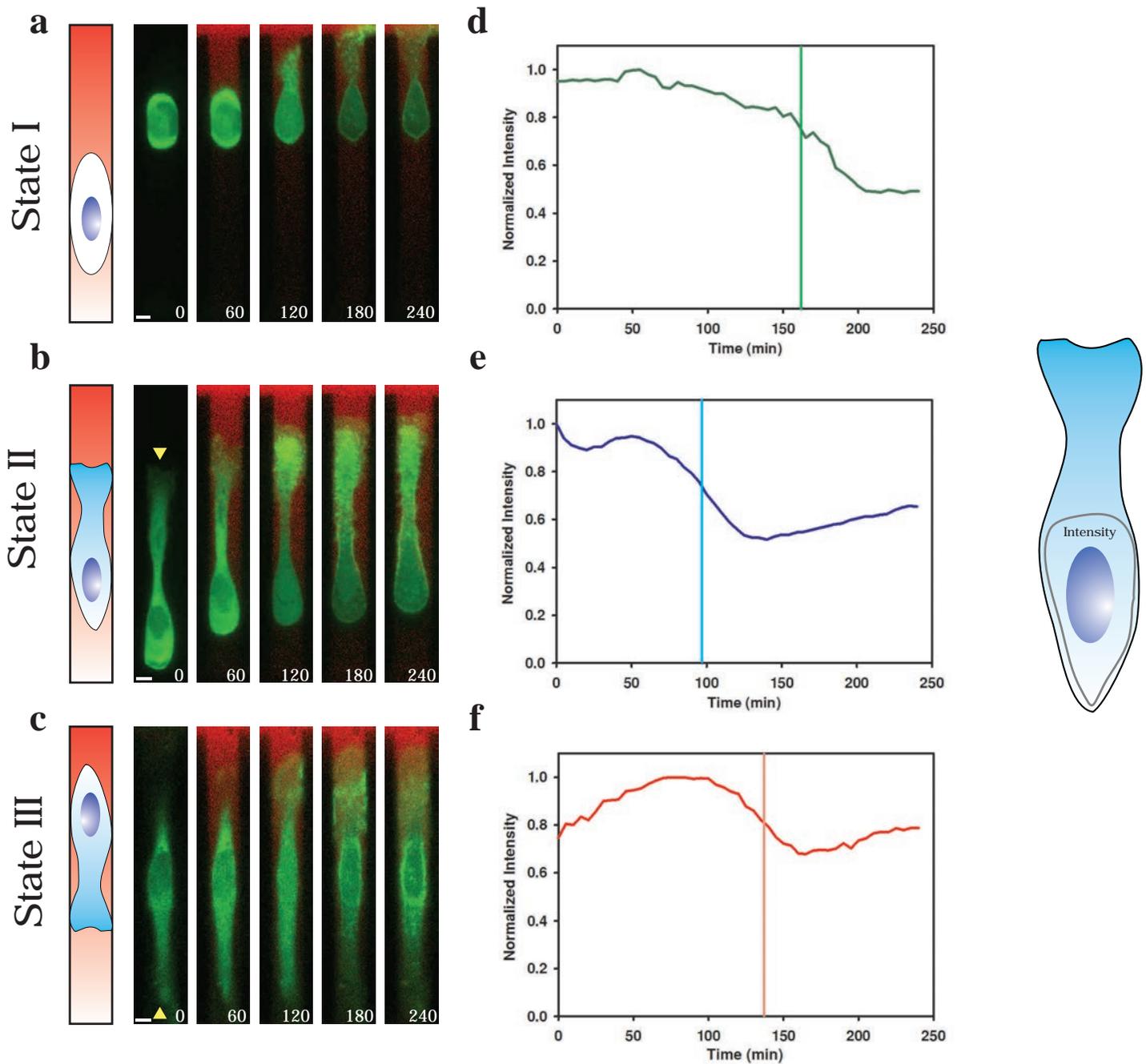

Supplementary Figure 9 | State II cells undergo the late polarization response faster than cells in other states. (a-c) The polarization response of cells in all three states experiencing the same concentration gradient and the same local concentration of rapamycin. Yellow arrowheads indicate the direction of the initial cell polarity. (d-f) Quantification of the average fluorescence intensity in cell bodies of the cells shown in (a-c) (see the schematic for the cell body definition). Drop lines indicate the late polarization times for each example. Note that the cell in state II reaches the late polarization phase faster than the cells in the other two states. Times are in minutes. Scale bars, 10 μm.

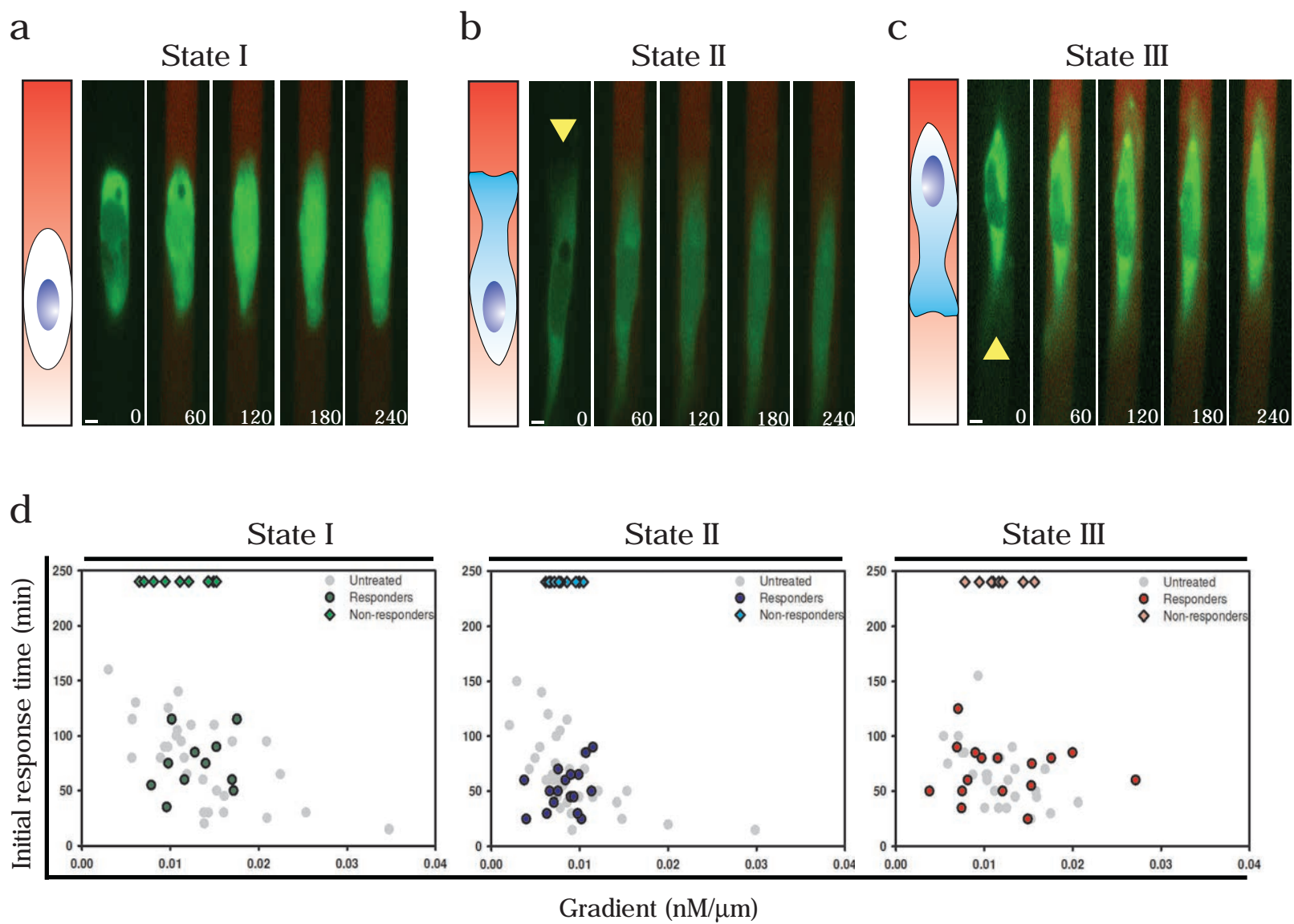

Supplementary Figure 10| Suppression of polarization responses with LY294002 treatment. (a-c) Time lapse imaging of sample cells selected from subpopulations in different states during LY294002 addition. LY294002 was included at concentration of 10 µm was included in both cell medium solutions used to generate the rapamycin gradient (Figure 1a). In all examples shown, cells fail to respond within the experimental time frame across all states. Yellow arrows indicate the direction of the initial cell polarity. Times are in minutes. Scale bars, 10 µm. (d) Initial response time vs. rapamycin gradient values in cells responding in the presence of LY294002. State I (green) n = 21, state II (blue) n = 31, and state III (red) n = 27. Diamonds represent non-responder cells while circles represent responding cells. Grey dots on each plot illustrate the initial response times seen for cells not exposed to LY294002 (Fig. 2).